\documentclass[reprint,aps,prx,showpacs,twocolumn]{revtex4-1}

\usepackage[utf8]{inputenc}
\usepackage{soul}
\usepackage{mathptmx}
\usepackage{mathtools}
\usepackage{amsmath} 
\usepackage{braket}
\usepackage{float}
\DeclarePairedDelimiter\abs{\lvert}{\rvert}%
\usepackage{lipsum}  
\usepackage{svg}
\let\vec\mathbf
\usepackage{mathptmx}
\usepackage{stackrel}
\usepackage{mathtools}
\usepackage{amsmath} 
\usepackage{amssymb}
\usepackage{braket}
\usepackage{float}
\usepackage{svg}
\let\vec\mathbf

\usepackage[sort&compress]{natbib}
\usepackage{mhchem}
\usepackage{changepage}
\usepackage{eucal}
\usepackage{float}
\usepackage{mathrsfs}
\usepackage{tikz}
\usepackage{mathrsfs}
\usepackage{todonotes}
\usepackage{tikz}
\usepackage{setspace}
\usepackage[nomargin]{fixme}
\usepackage[colorlinks= true, linkcolor=blue, citecolor=blue, urlcolor=blue]{hyperref}

\date{}

\begin{document}

\title{Continuous Gated First-Passage Processes}

\author{{\normalsize{}Yuval Scher$^{1,}$}
{\normalsize{}}}
\email{yuvalscher@mail.tau.ac.il}

\author{{\normalsize{}Aanjaneya Kumar$^{2,}$}
{\normalsize{}}}
\email{kumar.aanjaneya@students.iiserpune.ac.in}

\author{{\normalsize{}M. S. Santhanam$^{2,}$}
{\normalsize{}}}
\email{santh@iiserpune.ac.in}

\author{{\normalsize{}Shlomi Reuveni$^{1,}$}
{\normalsize{}}}
\email{shlomire@tauex.tau.ac.il}

\affiliation{\noindent \textit{$^{1}$School of Chemistry, Center for the Physics \& Chemistry of Living Systems, Ratner Institute for Single Molecule Chemistry, and the Sackler Center for Computational Molecular \& Materials Science, Tel Aviv University, 6997801, Tel Aviv, Israel \\
$^{2}$Department of Physics, Indian Institute of Science Education and Research, Dr. Homi Bhabha Road, Pune 411008, India.}}

\date{\today}

\begin{abstract}
Gated first-passage processes, where completion depends on both hitting a target and satisfying additional constraints, are prevalent across various fields. Despite their significance, analytical solutions to basic problems remain unknown, e.g., the detection time of a diffusing particle by a gated interval, disk, or sphere. In this paper, we elucidate the challenges posed by continuous gated first-passage processes and present a renewal framework to overcome them. This framework offers a unified approach for a wide range of problems, including those with single-point, half-line, and interval targets. The latter have so far evaded exact solutions. Our analysis reveals that solutions to gated problems can be obtained directly from the ungated dynamics. This, in turn, reveals universal properties and asymptotic behaviors, shedding light on cryptic intermediate-time regimes and refining the notion of high-crypticity for continuous-space gated processes. Moreover, we extend our formalism to higher dimensions, showcasing its versatility and applicability. Overall, this work provides valuable insights into the dynamics of continuous gated first-passage processes and offers analytical tools for studying them across diverse domains.
\end{abstract}

\maketitle

\section{Introduction} \label{Sec:1}
 
When does an observable that follows a random trajectory reach a certain target value for the first time? The answer to this question has proved to be valuable across many fields and disciplines, and an enormous amount of work has been dedicated to the study of this so called first-passage time \cite{redner2001guide,metzler2014first,klafter2011first,rudnick2004elements}. Yet, there are cases where it is not enough to simply arrive at the target. 

Consider for instance the binding and subsequent conversion of a substrate molecule to a product by enzymatic catalysis. Letting $X(t)$ denote the distance between the enzyme and its substrate, we see that if binding always occurs upon contact, the reaction time can be approximated by the first-passage time of $X(t)$ to zero  \cite{von1916drei,smoluchowski1918versuch,chandrasekhar1943stochastic,collins1949diffusion,calef1983diffusion}. This approximation neglects the possibility of substrate unbinding, and moreover assumes that the first-passage time to contact is long compared to the catalysis time. Yet, even when these assumptions hold, significant discrepancies between first-passage and reaction times may emerge due to stochastic transitions of the enzyme between a reactive state, at which it can bind the substrate, and a non-reactive state at which binding is not possible. 

\begin{figure}[t!]
\centering
\includegraphics[width=1\linewidth]{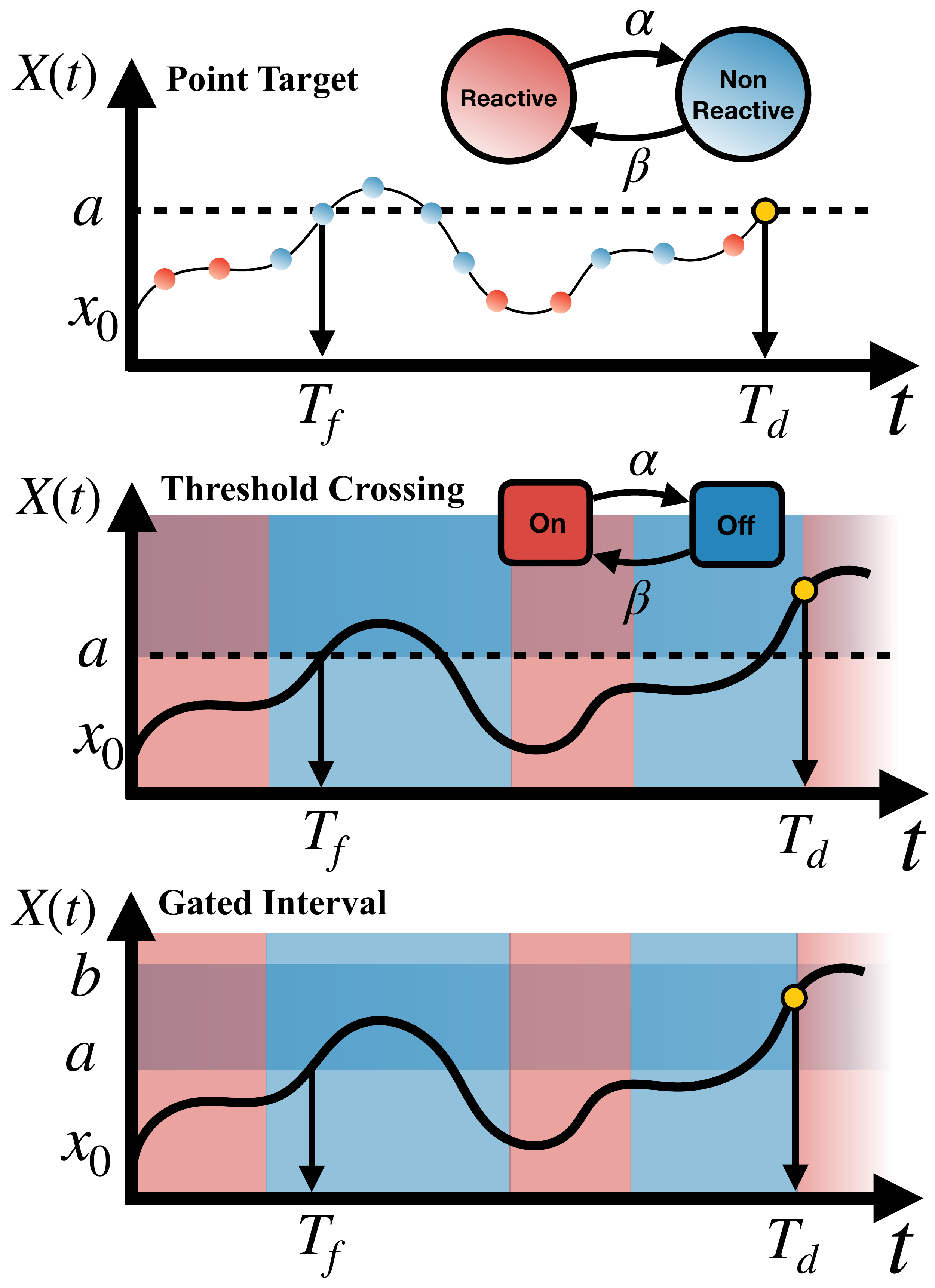}
\caption{Continuous gated first-passage processes. Two central examples of such processes are gated chemical reactions (top panel) and detection of threshold crossing by intermittent sensing (middle panel). Red represents the molecule being in the reactive state, or respectively the sensor being on. Blue represents the molecule being in the non-reactive state, or respectively the sensor being off. The corresponding first passage times of these processes are denoted by $T_f$, while the reaction/detection times are denoted by $T_d$. The point target (top) and threshold crossing (middle) scenarios can both be seen as special cases of the gated interval problem (bottom).}
\label{fig1}
\end{figure}

In the scenario described above, the enzyme acts as a ``gate" that has to be open to admit the substrate. This observation gave birth to the term ``gated reactions", which was originally introduced in the work of Perutz and Mathews on ligand binding by methaemoglobin \cite{perutz1966x}. Kinetic study of gated reactions in the bulk was initiated in the seminal works of McCammon, Northrup, Szabo, et al. \cite{mccammon1981gated,northrup1982rate,szabo1982stochastically}, and extensive research followed \cite{kim1992theory,zhou1996theory,berezhkovskii1997smoluchowski,re1996survival,makhnovskii1998stochastic,bandyopadhyay2000theoretical,benichou2000kinetics,bressloff2015stochastically,bressloff2016stochastic}. Single-particle models and approaches were also considered \cite{budde1995transient,caceres1995theory,spouge1996single,sheu1997survival,sheu1999first,bressloff2015escape,godec2017first,shin2018molecular,mercado2019first,mercado2021first,mercado2021search,scher2020unifying,scher2020gated,kumar2022inference}. Clearly, many of the results obtained there go beyond the context of chemical reactions and are also applicable elsewhere. An illustration of a gated reaction, i.e., a process $X(t)$ that ends upon reaching a point target in the reactive state, is given in the top panel of Fig.~\ref{fig1}.

Another particularly important instance where gating arises naturally is in the context of partially observable dynamical systems \cite{kumar2021first}. Suppose we are monitoring an observable of interest $X(t)$, which can denote diverse time-series ranging from the price of a commodity to the number of infected individuals in an epidemic. One is then often interested in the time it takes $X(t)$ to cross a threshold for the first time. Indeed, this first-passage time plays a significant role in various contexts, e.g., cell biology \cite{robert2015size,si2019mechanistic,rijal2020protein}, finance  \cite{valenti2007hitting,zhang2009first,wergen2011record,chicheportiche2013some,sabir2014record}, climate studies \cite{hoyt1981weather,schmittmann1999weather,redner2006role}, transport phenomena  \cite{barre2015generalized,barre2015cascading,barre2018stochastic,kishore2011extreme,kumar2020extreme}, transition path times on a reaction coordinate \cite{makarov2022effect,berezhkovskii2024significance}, and bio-chemical reactions  \cite{daoduc2010threshold,grebenkov2017first,lawley2019first,grebenkov2022reversible,grebenkov2022first,grebenkov2020surface,grebenkov2020paradigm}. It also lies at the heart of extreme-value \cite{kotz2000extreme,majumdar2020extreme,hartich2019extreme,sabhapandit2019extremes} and record statistics  \cite{chandler1952distribution,godreche2017record,majumdar2008universal,schehr2013exact,wergen2013records}, which aim to describe the properties of rare events occuring in dynamical systems \cite{malik2020rare}. 

It is oftentimes the case that $X(t)$ cannot be observed continuously at all times, and its value is thus known only at intermittent time intervals or points. The reason for this observed intermittency can range from sensor deficiencies and energy costs of continuous observations \cite{mitcheson2008energy,balsamo2014hibernus,hester2017future}, to simple lack of data in certain time windows \cite{rubin1976inference,kreindler2006effects}. Cause aside, the resulting outcome is the situation depicted in the middle panel of Fig.~\ref{fig1} (bottom), where the blue and red backgrounds denote the times when the notional sensor used to monitor the observable is on or off respectively. Akin to gated reaction, in situations pertaining to partial observations, the first-passage time $T_f$ can be missed and the relevant observable is the detection time $T_d$. The latter denotes the first time $X(t)$ is ``observed" above the pre-defined threshold \cite{kumar2021first}.

Driven by applications to the kinetics of chemical reactions, the work published on gating to date has mainly focused on understanding reaction time statistics for single-point gated targets. In comparison, the discussion of threshold-crossing phenomena under intermittent sensing is much more recent. Nevertheless, the chemical reaction perspective and the partially observable dynamical systems perspective, both fall under the class of gated-first passage processes. Thus, the terms ``reaction time" and ``detection time" are interchangeable and their use is context dependent. Moreover, both the single-point and threshold crossing problems can be seen as limits of a problem where an interval $[a,b]$ serves as the gated target (Fig.~\ref{fig1}, bottom). Indeed, the single-point target is obtained by taking the limit of $b \to a$, and the threshold crossing problem by taking $b \to \infty$. 

Surprisingly, despite the plethora of applications, the detection time of a particle inside a one-dimensional gated interval remains unknown. In fact, only very recently the analogous problem for simple diffusion with a partially absorbing region (radiative boundary conditions) was solved for spherically symmetric target intervals \cite{bressloff2022diffusion,schumm2021search,guerin2021universal}. This model was motivated by applications to biology, where the interval could represent a single reactive molecule, an intracellular compartment, or even a whole cell. Note that when the target is a single-point or embodied in the domain boundary, partial reactivity can be viewed as a limit of gating. Indeed, in the limit where the gating dynamics equilibrates much more rapidly than the spatial process, one can assume constant reactivity due to the separation of time scales. However, when the target is extended, e.g, an entire interval, within which the particle can diffuse (as is the case when the target has a permeable or semi-permeable boundary), partial reactivity is \textit{not} a limit of gating.

A notable general approach to gated reactions in continuous time and space is that of Szabo, Spouge, Lamm and Weiss (SSLW) \cite{spouge1996single,szabo1984localized}.  Their approach results in a general relation between the propagator of a Markovian gated process and its corresponding conserved spatial propagator (namely, in the absence of reaction/detection). From this relation, an expression for the detection time can be derived. Unfortunately, it seems that this valuable method was not used in subsequent literature on gated reactions, perhaps because it is given in the most general form possible. We demonstrate how the general equation of SSLW can be simplified to elegantly solve, e.g., the single-point target problem in Fig.~\ref{fig1} (top). Yet, we find that application of the same method to problems with finite-sized gated targets fails to yield a closed-form solution, and instead results in a complicated integral equation. Thus, to date, this fundamental problem remained unsolved.

To overcome the gated interval problem we hereby develop a renewal framework that allows us to solve for the first detection time of gated first-passage processes that undergo Markovian evolution in continuous space and time, e.g., diffusion. Our  approach and solution to the problem provide valuable insight into universal properties of gated first-passage processes. It also readily yields results obtained for the limiting cases of gated reactions with point-like targets \cite{mercado2019first}, and threshold crossing under intermittent sensing \cite{kumar2021first}, thus unifying treatment overall.

There has been a surge of interest in developing renewal frameworks to study universal features of stochastic processes with elements such as gating, \cite{scher2020unifying,scher2020gated,kumar2022inference,kumar2021first}, stickiness \cite{scher2023escape,grebenkov2023diffusion,grebenkov2021reversible,grebenkov2022first}, relocations \cite{eliazar2007searching} and resetting \cite{evans2011diffusion,evans2020stochastic,renewal2,renewal3,renewal4,renewal5,renewal6,renewal7,renewal9,renewal11,renewal13,kumar2023universal,renewal15,renewal16,renewal17,bonomo2021first}. In particular, recently this approach has seen great success in the study of gated reactions on networks \cite{scher2020unifying,scher2020gated}, which has opened the possibility of studying generalizations beyond Markovian gated reactions. These recent developments rely on the identification of the \emph{first-return time} of a particle back to its target as an instance of renewal, which in turn allows one to obtain closed form solutions and also gain insight that was previously overlooked in the literature. However, the story is very different when one thinks about continuous space, where the notion of first-return may be ill-defined and the treatment must hence be different from the one given for gating in discrete-space. In this paper, we will carefully explain the source of this difficulty and its implications. Next, we will provide a remedy that allows one to circumvent the problem altogether and make analytical progress. 

Apart from allowing solution to the aforementioned prototypical problem of the gated interval, we will explain how our formalism can be used as a tool for the analysis of other Markovian gated processes, and how it reveals universal features of this wide set of processes. Similar to discrete-space \cite{scher2020unifying,scher2020gated}, here too we provide formulas for the Laplace transform of the gated reaction time distribution and its mean. In cases where the mean reaction time diverges, we prove that the long-time asymptotics of the gated problem is inherited from its ungated counterpart, where only the pre-factor of the power-law tail changes. Interestingly, there is more to the story as gating can lead to a cryptic intermediate-time behaviour that precedes the long-time behaviour. As this intermediate-time behaviour may span over many decades, it can become  dominant as longer times may be too long to be observed.

Observation of cryptic intermediate-time behaviour starts with Mercado-Vásquez and Boyer who presented a comprehensive analysis of the gated reaction time of a freely diffusing particle with a single-point target (Fig.~\ref{fig1}, top) \cite{mercado2019first}. They identified conditions under which a transient regime of slower $\sim t^{-1/2}$ power-law decay emerges prior to the terminal $\sim t^{-3/2}$ asymptotics that governs the reaction time distribution in this problem. This behaviour was deemed cryptic, and gated diffusion was defined to be highly-cryptic if it spends most of its time in the non-reactive state. Shortly after, it was shown that under this high-crypticity condition a slower power-law decay at intermediate times is a universal feature displayed by a wide set of gated processes in discrete space and time \cite{scher2020gated}. Here, we extend this analysis to continuous-time. We find that conclusions do not generalize straightforwardly, and that an additional condition is required to produce the slower transient regime.

Simply put, the time spent in the non-reactive state before transitioning to the reactive state must be considerably larger than the time spent on the target upon arrival. Since Mercado-Vásquez and Boyer have explored the limiting case of a single-point target, this requirement was vacuously fulfilled. Indeed, when the target is of measure zero, the particle spends no time on the target and can react with it only via crossing it while being in the reactive state. Thus, in such cases, high-crypticity only cares about the equilibrium occupancies of the reactive and non-reactive states --- the rates themselves can be arbitrarily large, as long as the rate to become reactive is much slower than its converse. Clearly, this is not true for targets of finite size, where high transition rates ensure reaction (detection) upon arriving to the target.  This realization calls for a refined definition of high-crypticity for gated processes in continuous space and time. We also show that this refinement is relevant, and should be accounted for, when considering continuous time gated processes on networks.

Our paper is structured as follows. In Sec. \ref{Sec:2} we review the main results of the SSLW approach, while providing insight of our own. Markedly, we demonstrate how this approach can be used to elegantly solve the problem illustrated in Fig.~\ref{fig1} (top), which corresponds to a gated point target. Yet, In Sec. \ref{Sec:3}, we show that the application of the SSLW formalism to the detection of threshold crossing events under intermittent sensing (Fig.~\ref{fig1}, middle), leads to an integral equation which does not lend itself to solution by methods familiar to us.

In Sec. \ref{Sec:4}, we develop a renewal based approach to continuous gated processes. This approach allows us to elegantly bypass the limitations of the SSLW formalism. Instead of solving an integral equation, our formalism reduces the problem to that of solving a system of \emph{linear} equations. The solution allows us to represent the detection time density, for a  gated interval of arbitrary size, in terms of purely ungated observables. Furthermore, we then extract universal relations and demonstrate that these apply to a large set of gated processes. 

In Sec. \ref{Sec:5}, we use our formalism to solve a set of gated diffusion problems. We start, in Sec. \ref{sub:free}, with the fundamental problem of free diffusion to a gated interval-target (single-point target and threshold crossing included). We go on to study the effect of confinement, in Sec. \ref{sub:confinement}, where we solve for the case in which diffusion is restricted to an interval or slab. Confinement renders all moments of the detection time finite, but gating strongly affects the shape of the detection time distribution which need not be exponential. Yet, for the special case of high-crypticity, we observe that the distribution is well-described by a single exponent that sets  the mean detection time. In Sec. \ref{Sec:6}, we solve the problem of diffusion to a gated interval-target on a semi-infinite line with drift towards a reflecting boundary at zero. Similar to confined diffusion, this serves as an example of a process with finite moments of the ungated first-passage time distribution. Yet, we show that its gated counterpart can exhibit novel and remarkably different features. In particular, we demonstrate and discuss the presence of a non-monotonic dependence of the detection time on the drift velocity.

In Sec. \ref{Sec:7}, we further exemplify the power of our formalism to problems in higher dimensions. We generalize from a gated interval to a gated spherical target centered at the origin. To do so, we take advantage of the mapping between the d-dimensional Fokker-Planck equation for the distance of a freely diffusing particle from the origin (Bessel process) and one-dimensional diffusion on the semi-infinite line in a logarithmic potential. Under this mapping, free diffusion to a gated spherical target of radius $b$ can be mapped onto diffusion in a logarithmic potential to a gated target interval $[0,b]$.

We conclude this paper in Sec. \ref{Sec:8} with a summary and outlook.


\section{The Szabo, Spouge, Lamm and Weiss (SSLW) Approach to Gated Reactions}\label{Sec:2}

In this section, we revisit the SSLW approach \cite{spouge1996single,szabo1984localized} to the kinetics of gated reactions with the intention of reviewing its salient features, illustrating the power of this approach in the case of a gated point target. In the next section we will discuss the subsequent roadblocks in the generalization of this approach towards a continuum interval target.

We start by considering a system which can be characterized by keeping track of two variables: $\mathbf{r} \in V$, denoting the spatial coordinates of a particle in search of a gated target, and $\mathbf{\sigma} \in \Omega_Q$, denoting the internal state of the system. Thus, at any given time, the state of the system can be represented as $\mathbf{x}=(\mathbf{r},\mathbf{\sigma})$, and we assume that $\mathbf{x}$ undergoes Markovian evolution in the state space $\Omega=V \times \Omega_{Q}$, where $V$ is the spatial domain and $\Omega_{Q}$ is the set of possible gating states. 

The Markovian assumption allows us to write a mass-balance equation that expresses the propagator, $p\left(\mathbf{x}, t \mid \mathbf{x}_{0}\right)$, of the gated problem in terms of the conserved propagator, $G\left(\mathbf{x}, t \mid \mathbf{x}_{0}\right)$, for the corresponding problem with gating dynamics, but in the absence of an absorbing target. This reads 
\begin{equation}
\begin{aligned} \label{mass-balance}
   p\left(\mathbf{x}, t \mid \mathbf{x}_{0}\right)=& G\left(\mathbf{x}, t \mid \mathbf{x}_{0}\right)-\int_{0}^{t} d t^{\prime} \int_{\Omega} d \mathbf{X} \\
   & \times G(\mathbf{x}, t-t^{\prime} \mid \mathbf{X}) c(\mathbf{X}) p\left(\mathbf{X}, t^{\prime} \mid \mathbf{x}_{0}\right),
\end{aligned}
\end{equation}
where the function $c(\mathbf{X})$ accounts for the reactivity at point $\mathbf{X}$, where for all points but the target points we set $c(\mathbf{X})=0$. For a target point we set $c(\mathbf{X})=\kappa_{i} \delta\left(\mathbf{X}-\mathbf{a}_{i}\right)$, where $\mathbf{a}_{i} \in \Omega$ denotes a target at location $a_{i}$, which is  reactive for a certain state in $\Omega_Q$ with reactivity $\kappa_{i}$. If the target is completely absorbing we take the limit $\boldsymbol{\kappa}_{i} \to \infty$. Otherwise, we have a gated partially-reactive target with a reaction coefficient ${\kappa}_{i}$ on top of the gating rate.

For the case of $M$ point targets $c(\mathbf{X})=\sum_{i=1}^{M} \boldsymbol{\kappa}_{i} \delta\left(\mathbf{X}-\mathbf{a}_{i}\right)$, and Eq. (\ref{mass-balance}) boils down to
\begin{align} \label{mass-balance-points}
    p\left(\mathbf{x}, t \mid \mathbf{x}_{0}\right)=& G\left(\mathbf{x}, t \mid \mathbf{x}_{0}\right) \\
    & -\int_{0}^{t} d t^{\prime} \sum_{i=1}^{M} G\left(\mathbf{x}, t-t^{\prime} \mid \mathbf{a}_{i}\right) \boldsymbol{\kappa}_{i} p\left(\mathbf{a}_{i}, t^{\prime} \mid \mathbf{x}_{0}\right). \nonumber
\end{align}
By Laplace transforming $\tilde{f}(s) = \int_0^{\infty} d t^{\prime} f(t^{\prime}) e^{-st'}$, and setting $\mathbf{x}=\mathbf{a}_{i}$, one obtains a set of $M$ linear equations for $\tilde{p}\left(\mathbf{a}_{i}, s \mid \mathbf{x}_{0}\right)$. Solving, and substituting back into Eq. (\ref{mass-balance-points}), we find  
\begin{equation}
\begin{aligned} \label{Laplace-mass-balance-points}
    \tilde{p}\left(\mathbf{x}, s \mid \mathbf{x}_{0}\right)=& \tilde{G}\left(\mathbf{x}, s \mid \mathbf{x}_{0}\right)-\left[\boldsymbol{\kappa}_{j} \tilde{G}\left(\mathbf{x}, s \mid \mathbf{a}_{j}\right)\right] \\
    & \times\left[\delta_{i j}+\kappa_{j} \tilde{G}\left(\mathbf{a}_{i}, s \mid \mathbf{a}_{j}\right)\right]^{-1}\left[\tilde{G}\left(\mathbf{a}_{i}, s \mid \mathbf{x}_{0}\right)\right],
\end{aligned}
\end{equation}
where the final three factors are, respectively, $1 \times M$, $M \times M$, and $M \times 1$ matrices, and $\delta_{i j}=1$ for $i=j$ and zero otherwise \cite{szabo1984localized}. Note that the matrix multiplication of the last three terms gives a $1 \times 1$ term. We will refer to this equation and its analogues as the SSLW equation.

When considering a single point target, $M=1$, Eq. (\ref{Laplace-mass-balance-points}) simplifies considerably 
\begin{equation} \label{SSLW_single_radiative}
  \tilde{p}\left(\mathbf{x}, s \mid \mathbf{x}_{0}\right)=\tilde{G}\left(\mathbf{x}, s \mid \mathbf{x}_{0}\right)-\frac{\kappa \tilde{G}\left(\mathbf{x}, s \mid \mathbf{a}_{+}\right) \tilde{G}\left(\mathbf{a}_{+}, s \mid \mathbf{x}_{0}\right)}{1+\kappa \tilde{G}\left(\mathbf{a}_{+}, s \mid \mathbf{a}_{+}\right)},
\end{equation}
where $\mathbf{a}_{+}=(+, \mathbf{a})$ denotes a single trap at $\mathbf{r}=\mathbf{a}$, which can trap only when the system is in a single active state $\sigma=(+)$. The survival probability for a gated particle with initial state $\mathbf{x}_{0}$ is given by
\begin{equation} \label{survival_integral}
    S\left(t \mid \mathbf{x}_{0}\right)=\int_{\Omega} p\left(\mathbf{X}, t \mid \mathbf{x}_{0}\right) d \mathbf{X}.
\end{equation}
Laplace transforming and integrating Eq. (\ref{survival_integral}) one gets 
\begin{equation} \label{survival}
    \tilde{S}\left(s \mid \mathbf{x}_{0}\right)=s^{-1}\left[1-\frac{\kappa \tilde{G}\left(\mathbf{a}_{+}, s \mid \mathbf{x}_{0}\right)}{1+\kappa \tilde{G}\left(\mathbf{a}_{+}, s \mid \mathbf{a}_{+}\right)}\right].
\end{equation}

So far the main results from the work of Spouge, Szabo and Weiss on gated reactions that will be relevant for our discussion here \cite{spouge1996single}. In their paper, the focus then shifts to cases in which an equilibrium probability distribution for the conserved propagator exists or can be assumed by imposing a uniform initial condition, which is especially valuable for macroscopic kinetic analysis. 

Note, however, that in deriving Eq. (\ref{SSLW_single_radiative}) the only assumption was Markovian evolution in the joint state space $\Omega=V \times \Omega_{Q}$. In practice, when modeling gated reactions, a two-state Markovian internal dynamics which is decoupled from the spatial process is usually assumed. Thus, in the next subsection we explicitly make this assumption and show how Eq. (\ref{SSLW_single_radiative}) consequently simplifies. Similarly, we will also assume that $\kappa \to \infty$, thus putting complications coming from partial absorption aside. Clearly, other simplifications of Eq. (\ref{SSLW_single_radiative}) can be made and analyzed.

\subsection{The SSLW equation for Markovian (Poissonian) gating of an absorbing single-point target}

Assuming that the reactive state is completely absorbing, we take the limit $\kappa \to \infty$ in Eq. (\ref{SSLW_single_radiative}) and obtain 
\begin{equation} \label{SSLW_single_absorbing}
  \tilde{p}\left(\mathbf{x}, s \mid \mathbf{x}_{0}\right)=\tilde{G}\left(\mathbf{x}, s \mid \mathbf{x}_{0}\right)-\frac{\tilde{G}\left(\mathbf{x}, s \mid \mathbf{a}_{+}\right) \tilde{G}\left(\mathbf{a}_{+}, s \mid \mathbf{x}_{0}\right)}{\tilde{G}\left(\mathbf{a}_{+}, s \mid \mathbf{a}_{+}\right)}.
\end{equation}
In the same manner, taking the limit $\kappa \to \infty$ in Eq. (\ref{survival}) we obtain
\begin{equation} \label{survival_absorbing}
    \tilde{S}\left(s \mid \mathbf{x}_{0}\right)=s^{-1}\left[1-\frac{ \tilde{G}\left(\mathbf{a}_{+}, s \mid \mathbf{x}_{0}\right)}{ \tilde{G}\left(\mathbf{a}_{+}, s \mid \mathbf{a}_{+}\right)}\right].
\end{equation}

We use $T_d(\mathbf{x}_{0})$ to denote the random detection time of a particle with initial state $\mathbf{x}_{0}$, and denote the probability density of this random variable by $f_d\left(t \mid \mathbf{x}_{0}\right)$. In cases where we explicitly define a random variable $T$, we denote the Laplace transform of its distribution $f(t)$ by $\tilde{T}(s) = \int_0^\infty dt^{\prime} f(t^{\prime})e^{-s t^{\prime}}$. Recalling that $S(t \mid \mathbf{x}_{0})=1-\int_{0}^{t} d t^{\prime} f_d\left(t^{\prime} \mid \mathbf{x}_{0}\right) $, Laplace transforming and using Eq. (\ref{survival_absorbing}), we obtain
\begin{equation} \label{FP_propagtor_relation}
\tilde{T}_d(s \mid \mathbf{x}_{0})=\frac{ \tilde{G}(\mathbf{a}_{+}, s \mid \mathbf{x}_{0})}{ \tilde{G}(\mathbf{a}_{+}, s \mid \mathbf{a}_{+})},
\end{equation}
which generalizes a well-known relation for first-passage times (see, for example, Eq. (2.9) in Ref. \cite{klafter2011first}) to detection times. 

A closer look at Eq. (\ref{FP_propagtor_relation}) allows us to write an integral equation satisfied by the gated reaction time distribution to reach a point target as follows
\begin{equation} \label{FP_prop_reltation_interpret}
    G(\mathbf{a}_{+}, t \mid \mathbf{x}_{0}) =\int_0^t f_d(t' \mid \mathbf{x}_{0})  G(\mathbf{a}_{+}, t-t' \mid \mathbf{a}_{+}) ~ dt'.
\end{equation}

\noindent This yields the following interpretation: Each trajectory in which the particle is in state $\mathbf{a}_{+}$ at time $t$, can be decomposed in two parts where: (i) starting from $\mathbf{x}_{0}$, the particle is first detected at the $\mathbf{a}_{+}$ state at a time $t'<t$, and (ii) starting from $\mathbf{a}_{+}$, the particle continues evolving for an additional $t-t'$ time units, such that at time $t$ it is back at $\mathbf{a}_{+}$ again. Integrating over all intermediate values of $0<t'<t$, we recover Eq.~\eqref{FP_prop_reltation_interpret}.

To proceed, let us also assume that the internal and spatial dynamics are decoupled, such that
\begin{equation} \label{decoupled}
    G\left(\mathbf{x}, t \mid \mathbf{x}_{0}\right) = C\left(\mathbf{r}, t \mid \mathbf{r}_{0}\right) Q\left(\mathbf{\sigma}, t \mid \mathbf{\sigma}_{0}\right).
\end{equation}
Thus, by definition, 
\begin{equation} \label{decoupled_and_Laplaced}
    \tilde{G}\left(\mathbf{x}, s \mid \mathbf{x}_{0}\right) = \int_{0}^{\infty} C \left(\mathbf{r}, t \mid \mathbf{r}_{0}\right) {Q}\left(\mathbf{\sigma}, t \mid \mathbf{\sigma}_{0}\right) e^{-st}dt.
\end{equation}
To further proceed, we consider two-state Markovian gating with $\sigma_{0} \in \{\text{R, NR}\}$, i.e., a reactive state and a non-reactive state. The transition rate from R to NR is denoted by $\alpha$ and the transition rate from NR to R is denoted by $\beta$. The internal dynamics propagator is then
\begin{equation} \label{internal_propagator}
  \begin{array}{ll}
   Q(\text{R},t \mid \text{NR}) =\pi_{\textrm{R}}(1-e^{-\lambda t}) ,     \\
   Q(\text{NR},t \mid \text{NR}) =\pi_{\textrm{NR}}+\pi_{\textrm{R}}e^{-\lambda t}  ,     \\
   Q(\text{R},t \mid \text{R}) =\pi_{\textrm{R}}+\pi_{\textrm{NR}}e^{-\lambda t},     \\
   Q(\text{NR},t \mid \text{R}) =\pi_{\textrm{NR}}(1-e^{-\lambda t}),     
  \end{array}
\end{equation}
where $\lambda = \alpha + \beta$ is the relaxation rate to equilibrium, and $\pi_{\textrm{R}}=\beta / \lambda$, $\pi_{\textrm{NR}}=\alpha / \lambda$ are the equilibrium occupancies of the reactive and non-reactive states.

Substituting Eq. (\ref{internal_propagator}) into Eq. (\ref{decoupled_and_Laplaced}) and utilizing Eq. (\ref{FP_propagtor_relation}), we obtain 
the Laplace transform of the detection time 
\begin{equation} \label{SSLW_used}
 \tilde{T}_d(s \mid \mathbf{x}_{0})= \frac{\pi_{\textrm{R}}\tilde{C}\left(\mathbf{a}, s \mid \mathbf{r}_{0}\right) \pm (1-\pi_{\sigma_{0}}) \tilde{C}\left(\mathbf{a}, s+\lambda \mid \mathbf{r}_{0}\right)  }{\pi_{\textrm{R}}\tilde{C}\left(\mathbf{a}, s \mid \mathbf{a}\right) + \pi_{\textrm{NR}} \tilde{C}\left(\mathbf{a}, s+\lambda \mid \mathbf{a}\right)} , 
\end{equation}
where we have a plus sign if $\sigma_{0}=\text{R}$ and a minus sign if $\sigma_{0}=\text{NR}$. Of course, this relation can be plugged back into Eq. (\ref{survival_absorbing}) to give the corresponding survival probability.  

We believe it is plausible that the SSLW approach to gated reactions has not been widely applied in subsequent literature because it takes a very general form that appears to be more difficult to apply than it actually is. However, it can be appreciated that the more specialized relation in Eq. (\ref{SSLW_used}) only requires plugging in the underlying spatial propagator of the specific problem at hand, and perhaps some algebraic simplifications. To demonstrate the strength of this ``plug and play" result, we briefly consider the two examples below.

\subsection{Free diffusion with a gated target at the origin}

Plugging the Laplace transform of the propagator for one-dimensional free-diffusion, $ \tilde{C}(x, s | x_{0}) =\frac{1}{2\sqrt{sD}} e^{-\sqrt{\frac{s}{D}} \abs{x-x_{0}}}$, into Eq. (\ref{SSLW_used}) we obtain the reaction time of a particle on the infinite line, with a gated target at $x=0$. This result checks with the result of Mercado-Vásquez and Boyer \cite{mercado2019first}. Note the minimal effort invested in obtaining the solution to this fundamental problem which has just recently been solved by direct methods.


\subsection{Diffusion on the semi-infinite line with a gated target at the origin}

Plugging in the Laplace transform of the propagator for one-dimensional free-diffusion with reflecting boundary at the origin (Appendix \ref{Appendix:A})
\begin{equation} 
 \tilde{C}(x, s | x_{0}) = \frac{1}{\sqrt{s D}}\begin{cases}
   e^{-\sqrt{\frac{s}{D}}x_{0}}\hspace{1pt}\text{cosh}(\sqrt{\frac{s}{D}}x),  \hspace{10pt} x < x_{0},\\
  \text{cosh}(\sqrt{\frac{s}{D}}x_{0})e^{-\sqrt{\frac{s}{D}}x}, \hspace{10pt}  x > x_{0},\\
 \end{cases}    
\end{equation}
into Eq. (\ref{SSLW_used}), we get the reaction time distribution for a particle on the semi-infinite line with a gated target at $x=0$. Note that in this scenario, if the target is non-reactive, the particle is reflected from the origin. Clearly one obtains the exact same reaction time distribution as in the infinite line example, as can be easily verified. The two cases differ, however, in their propagators which can be obtained by carefully plugging the appropriate spatial and internal propagators into Eq. (\ref{SSLW_single_absorbing}).


\subsection{Gated reactions on networks}

We mention in passing that in deriving Eq. (\ref{mass-balance-points}) it was not assumed at any point that the space is continuous, and in fact this equation can be used to analyze discrete-space processes as well, such as random walks on networks. In Ref. \cite{scher2020unifying} we developed a universal formalism for gated reactions on networks. In Appendix \ref{Appendix:B}, we show how Eq. (7) of Ref. \cite{scher2020unifying} can be reproduced via a different route that starts with Eq. (\ref{mass-balance-points}). However, while the SSLW approach assumes Markovianity of both the spatial and internal processes, the renewal approach of Ref. \cite{scher2020unifying} allows for non-Markovian spatial processes, and thus extends the range of validity from Markovian random walks on networks to continuous-time random walks (CTRWs) on networks.

\section{Beyond localized targets}\label{Sec:3}
As demonstrated in the previous section, the SSLW formalism provides an elegant way for studying the first detection time density in the special case of a gated point target. Another important application of gated first-passage processes emerges in the study of threshold crossing events, where one is interested in computing the statistics of the time when a stochastic time-series is observed to be above a certain pre-defined threshold for the first time. 
In the previous section we have simplified the mass balance Eq. (\ref{mass-balance}) by utilizing the mathematical properties of the delta-functions representing a set of point targets. Unfortunately, we do not have in our arsenal such a trick for continuous targets. We must, therefore, resort back to Eq. (\ref{mass-balance}) and reassess. Laplace transforming gives 
\begin{equation}
\begin{aligned}
   \tilde{p}\left(\mathbf{x}, s \mid  \mathbf{x}_{0}\right) = & \tilde{G}\left(\mathbf{x}, s \mid \mathbf{x}_{0}\right) \\&- \int_{\Omega} d \mathbf{X} 
   \tilde{G}(\mathbf{x}, s \mid \mathbf{X}) c(\mathbf{X}) \tilde{p}\left(\mathbf{X}, s \mid \mathbf{x}_{0}\right).
\end{aligned}
\end{equation}





\noindent Thus, the SSLW formalism yields an integral equation for $\tilde{p}\left(\mathbf{x}, s \mid \mathbf{x}_{0}\right)$, but this does not seem to be immediately solvable. Moreover, even for the emblematic case of free diffusion under independent Markovian gating, where translation invariance implies that $G(\mathbf{x}, t \mid \mathbf{X}) = C(x-X,t|0) Q(\sigma,t|\sigma_0)$, the simplifications that arise do not lend themselves towards a closed-form solution for $\tilde{p}\left(\mathbf{x}, s \mid \mathbf{x}_{0}\right)$.

To progress further, a seemingly promising direction is to integrate both sides of the above equation over all values of $\mathbf{x}$, like in Eq.~\eqref{survival_integral}, to obtain an equation containing the survival probability. Doing so, we get 
\begin{equation}
\begin{aligned}
   \tilde{S}\left( s \mid \mathbf{x}_{0}\right) = \frac{1}{s}\bigg(1 - \int_{\Omega} d \mathbf{X} ~
      \tilde{p}\left(\mathbf{X}, s \mid \mathbf{x}_{0}\right)c(\mathbf{X})\bigg),
\end{aligned}
\end{equation}
which, despite making the equation more compact, does not bring us any closer to an exact solution for the survival probability, or the detection time density. 

We note that the challenge in obtaining the exact detection time density is a generic feature of extended continuous targets, and is not specific to the threshold crossing problem, where the gated target extends to infinity. In fact, this mathematical difficulty that one faces when going beyond a collection of point targets, and instead considering a continuous target interval was specifically noted in Ref. \cite{szabo1984localized}, and the possibility of developing a perturbation scheme for small values of reactivity was discussed \cite{weiss1984perturbation}. However, the limited applicability and accuracy of the approximate schemes subsequently developed warrants a new approach to the problem. In the next section, we introduce a renewal framework that allows us to obtain an exact solution to this challenging problem.

\section{Renewal Apparoach to Continuous Gated First-Passage Processes}\label{Sec:4}

\subsection{The renewal equations}

Consider a continuous one-dimensional stochastic process $X_{x_0}(t)$ that undergoes Markovian evolution, with $X_{x_0}(0)=x_0$, and a gated interval $[a,b]$, that stochastically switches between reactive (R) and non-reactive (NR) states. We define the random variable $T_d(x_0,\sigma_0)$, with $\sigma_{0} \in \{\text{R, NR}\}$, to be the detection time starting from the composite state $(x_0,\sigma_0)$. Namely, $T_d(x_0,\sigma_0)$ is the first time the process $X_{x_0}(t)$ is in the interval $[a,b]$, \emph{while} the gate is in its reactive state $R$. Let us define the shorthand notation 
\begin{equation}  \label{rho}
\rho = 
    \begin{cases}
    a, &\text{if~~} x_0 \le a,\\
    b, &\text{if~~} x_0 \ge b. \\
    \end{cases}
\end{equation}
We can express $T_d(x_0,\sigma_0)$ recursively in the following manner
\begin{equation}  \label{renewal_interval1}
T_d(x_0,\sigma_0)= T_f(\rho \mid x_0) +
    \begin{cases}
      0, &\text{if~~} \sigma_{T_f(\rho|x_0)}=\text{R}\\
       T_d(\rho,\text{NR}), &\text{otherwise},
    \end{cases}
\end{equation}
where $T_f(\rho \mid x_0)$ is a random variable that denotes the ungated first-passage time to $\rho$, starting from $x_0$, and $\sigma_{T_f(\rho|x_0)}$ denotes the state of the gate at this time.

In turn, the recursive relation satisfied by $T_d(\rho,\text{NR})$ follows a different logic: Instead of analysing the renewal of the process when the \textit{first return} to the gated interval happens, we consider the event where the gating state switches from the state NR to R, for the first time. This allows us to write  
\begin{equation}  \label{renewal_interval2}
T_d(\rho,\text{NR})= W_\beta +
    \begin{cases}
      0, &\text{if~~} X_{\rho}(W_\beta) \in [a,b], \\
       T_d(y,\text{R}), &\text{if~~} X_{\rho}(W_\beta) = y \in (-\infty, a),\\
       T_d(y,\text{R}), &\text{if~~} X_{\rho}(W_\beta) = y \in (b, \infty),
    \end{cases}
\end{equation}
where $W_\beta$ is the random variable that denotes the time taken to switch from state NR to R.

The structure of Eq. (\ref{renewal_interval2}) embodies the crucial difference between continuous and discrete space gated processes. In continuous space, the first-return time is ill-defined. In fact, a diffusing particle returns infinitely many times to the target
within an infinitely short time period \cite{grebenkov2020surface}. This presents a significant challenge, since in analyzing discrete space gated processes we have identified returns to the target as renewal moments. This, in turn, allowed us to write the first detection time solely in terms of the ungated first return time and transitions rates between the reactive and non-reactive states \cite{scher2020unifying,scher2020gated}.

To circumvent the ill-defined first-return time, we break the subsequent trajectory in Eq. (\ref{renewal_interval2}) into two: (i) First we let the process spatially evolve as if there is no target for the time $W_\beta$ it takes the particle to become reactive. During this time the system is non-reactive, and so it is safe to ignore the target. The probability to find the particle at some point $y$ after $W_\beta$ is given by the conserved spatial propagator $C(y, W_\beta \mid \rho)$, i.e., the propagator for the corresponding problem in the absence of reaction/detection. (ii) The particle then continues from position $y$ and state R. If $y$ happens to fall inside the target interval, we are done. Otherwise, the particle is either above or below the target. The detection time from that state is respectively given by plugging $x_0=y$ in Eq. (\ref{renewal_interval1}) and the corresponding value of $\rho$ according to Eq. (\ref{rho}).

Two things are immediately apparent. First, for this trick to work, we require Markovianity of the spatial process -- the exact trajectory that the particle has taken to reach the point $y$ in stage (i) is irrelevant when stage (ii) begins. Second, to know the statistics of the point $y$, knowledge of the corresponding conserved spatial propagator is required. In turn, there is no way to express the first detection time solely in terms of the ungated first-passage time and transitions rates, and the conserved spatial propagator is carried into this relation. Note that this unavoidable addition is also apparent in the SSLW formalising. For example, there is no way to completely get rid of the conserved spatial propagator in Eq. (\ref{SSLW_used}). These two realizations are in stark contrast to the analogue theory of discrete space gated processes which neither requires the spatial processes to be Markovian nor a knowledge of the propagator \cite{scher2020unifying,scher2020gated}. Nonetheless, the additional requirement in continuous space is a necessary price to pay for the solution of a much more complex gated problem.
\subsection{Mean detection time}\label{sub_mean_detection}

Let us assume that the mean detection time is finite, later on we will deal with cases of diverging mean.
\noindent Taking expectations of both sides of Eq. (\ref{renewal_interval1}) we obtain
\begin{equation} \label{mean1}
\Braket{T_d(x_0,\sigma_0)} = \Braket{T_f(\rho \mid x_0 )} + \Braket{I_f} \Braket{T_d(\rho, \text{NR})},
\end{equation}
where $I_f$ is an indicator random variable that receives the
value $1$ if the particle first arrived at $\rho$ in the nonreactive state and $0$ otherwise. In Eq. (\ref{mean1}) we have used the independence of $I_f$ and $T_d(\rho, \text{NR})$: $\Braket{I_f T_d(\rho, \text{NR})}=\Braket{I_f} \Braket{T_d(\rho, \text{NR})}$. For this, we have noted that $T_d(\rho, \text{NR})$ is the additional time it takes the reaction to complete in a scenario where the particle arrived at $\rho$ in the non-reactive state, i.e., conditioned on $I_f=1$. Thus, while $I_f$ determines if an additional time $T_d(\rho, \text{NR})$ should be added or not, it is uncorrelated with the duration of this time. The duration of $T_d(\rho, \text{NR})$ does not depend on whatever happened prior to arriving at the interval boundary. The expectation of the indicator function is
\begin{equation} \label{mean2}
\Braket{I_f} = \Braket{Q(\text{NR},T_f(\rho\mid x_0 ) \mid \sigma_0)}  .    
\end{equation}
Recalling Eq. (\ref{internal_propagator}), when $\sigma_0=\text{NR}$ we have
\begin{equation} \label{mean3}
\Braket{I_f (\sigma_0=\text{NR}) }=\pi_{\textrm{NR}}+\pi_{\textrm{R}}\tilde{T}_f(\rho,\lambda\mid x_0),
\end{equation}
where $\tilde{T}_f(\rho,\lambda\mid x_0)$ is the Laplace transform of $T_f(\rho\mid x_0)$ evaluated at $\lambda$.
Similarly, when $\sigma_0=\text{R}$, we have 
\begin{equation} \label{mean4}
\Braket{I_f (\sigma_0=\text{R})}=\pi_{\textrm{NR}}\Big[1-\tilde{T}_f(\rho,\lambda \mid x_0)\Big].   
\end{equation}
Equations (\ref{mean3}) or (\ref{mean4}) can in turn be plugged into Eq. (\ref{mean1}) in accordance with the initial condition $\sigma_0$. Note that if the initial state of the gating dynamics is the equilibrium occupancy probabilities, denoted here by $\sigma_0=\textrm{eq}$, we simply have
\begin{equation} \label{mean4}
\Braket{I_f (\sigma_0=\text{eq})}=\pi_{\textrm{NR}}.   
\end{equation}
Taking expectations of both sides of Eq. (\ref{renewal_interval2}) we obtain
\begin{align} \label{mean5}
\Braket{T_d(\rho, \text{NR})} & =\beta^{-1}  \\&+ \int^{\infty}_{-\infty} \tilde{\Phi}_{\rho}(\beta)  \Braket{T_d(y,\text{R})} \Big[\Theta_{-}(y)+\Theta_{+}(y)\Big] dy,  \nonumber
\end{align}
where $\tilde{\Phi}_{\rho}(z):=\beta \tilde{C}(y,z \mid \rho)$, such that $\tilde{C}(y,\beta \mid \rho)$ is the Laplace transform of $C(y, t \mid \rho)$ evaluated at $\beta$, and where $\Theta_{-}(y)$ is a step function that equals $1$ for all $y<a$ and $0$ otherwise, and $\Theta_{+}(y)$ equals $1$ for all $y>b$ and $0$ otherwise. In deriving Eq. (\ref{mean5}) we have used the independence of stages (i) and (ii) that were described below Eq. (\ref{renewal_interval2}), which requires Markovianity of the spatial propagator. Note that we have also used $\Braket{C(y,W_\beta \mid \rho)}=\beta \tilde{C}(y,\beta \mid \rho)=\tilde{\Phi}_{\rho}(\beta)$. 

We can now plug Eq. (\ref{mean1}) into Eq. (\ref{mean5}) while noting that $\Braket{T_d(\rho, \text{NR})}$ of Eq. (\ref{mean1}) is independent of $y$ in the integral of Eq. (\ref{mean5}), and so can be taken out of the integral. This gives
\begin{align} \label{mean6}
     \Braket{T_d(\rho, \text{NR})} =& \beta^{-1} + \tau_{\rho} 
     \\
     &+  p^{-}_{\rho} \Braket{T_d(a, \text{NR})} +  p^{+}_{\rho} \Braket{T_d(b, \text{NR})},  \nonumber
\end{align}
where we have defined
\begin{equation} \label{mean7}
\tau_{\rho} = \int^{\infty}_{-\infty} \tilde{\Phi}_{\rho}(\beta) \braket{T_f(\iota_{\pm}\mid y)}  \Big[\Theta_{-}(y)+\Theta_{+}(y)\Big] dy , 
\end{equation}
such that $\iota_- = a$ for $y<a$ and $\iota_+ = b$ for $y>b$, and
\begin{equation} \label{mean8}
  p^{\pm}_{\rho} =  \int^{\infty}_{-\infty} \tilde{\Phi}_{\rho}(\beta)
 \Braket{Q(\text{NR},T_f(\iota_\pm \mid y ) \mid \text{R})}   \Theta_{\pm}(y) dy ,
\end{equation}
where $\Braket{Q(\text{NR},T_f(\iota_{\pm}\mid y ) \mid \text{R})}=\pi_{\textrm{NR}}\Big[ 1  - \tilde{T}_f(\iota_{\pm},\lambda\mid y) \Big]$. Note that Eq.~(\ref{mean6}) is actually a shorthand notation for a system of two equations, for the two unknowns $\Braket{T_d(a, \text{NR})}$ and $\Braket{T_d(b,\text{NR})}$, which one gets by substituting the two possible values of $\rho=\{a,b\}$. 

Each term on the right-hand side of Eq.~(\ref{mean6}) gives us insight into the different mechanisms through which a detection event can take place. For instance, the first term $\beta^{-1}$ denotes the mean time taken for the particle to turn reactive (R), starting from the non-reactive state (NR). It is easy to see that the mean detection time $\Braket{T(\rho, \text{NR})}$ satisfies $\Braket{T(\rho, \text{NR})} \geq \beta^{-1}$ where the equality holds only in the extreme cases where the particle is always detected as soon as it turns reactive. However, in almost all practically relevant scenarios this will not be the case. Namely, there will be a non-zero probability for a particle that starts at the boundary of the gated interval to be found \emph{outside} the interval when it turns reactive. In this case, the additional time taken for detection is captured by the other three terms in Eq.~(\ref{mean6}). 

If detection did not happen when the particle turned reactive, it can happen later in two different ways. Suppose that when the particle turns reactive, it is at $y \notin [a,b]$. For a detection event to take place, the particle now has to reach the boundary nearest to it, starting from position $y$. If at the moment when the particle reaches the nearest boundary, it is found reactive, then it is detected right away. In Eq.~(\ref{mean6}), $\tau_{\rho}$ captures the weighted contribution of such events, starting from different values of $y$, to the mean detection time. However, if upon reaching the boundary closest to it, the particle is non-reactive, the dynamics is renewed and the mean additional time taken for detection is either $\Braket{T_d(a, \text{NR})}$ or $\Braket{T_d(b, \text{NR})}$, depending on whether $y$ was closer to boundary point $a$ or $b$. 

The identification of a renewal moment provides us with the needed closure that allows us to obtain the exact formula for the mean detection time. In particular, by solving the system of equations in~(\ref{mean6}) for the two unknowns $\Braket{T_d(a, \text{NR})}$ and $\Braket{T_d(b,\text{NR})}$, we obtain
\begin{equation}\label{mean9}
  \Braket{T_d(a, \text{NR})}    = \frac{(\beta^{-1}+\tau_a)(1-p^{+}_{b}) + p^{+}_{a} (\beta^{-1}+\tau_b)}{1 - p^{-}_{a} - p^{+}_{b} + p^{-}_{a} p^{+}_{b} - p^{-}_{b} p^{+}_{a} }, 
\end{equation} 
\noindent and
\begin{equation}\label{mean10}
\Braket{T_d(b, \text{NR})}    =
  \frac{(\beta^{-1}+\tau_b)(1-p^{-}_{a}) + p^{-}_{b} (\beta^{-1}+\tau_a)}{1 - p^{-}_{a} - p^{+}_{b} + p^{-}_{a} p^{+}_{b} - p^{-}_{b} p^{+}_{a} }. 
\end{equation} 

For the symmetric case in which the spatial dynamics and the boundary conditions to the left and the right of the target center are the same (e.g., diffusion on the infinite line), Eqs. (\ref{mean9}) and (\ref{mean10}) are equal and simplify considerably. Setting $\tau_a = \tau_b:=\tau$ and $p^{\pm}_{a} = p^{\pm}_{b}:=p^{\pm}$, we obtain
\begin{equation} \label{mean11}
  \Braket{T_d(\rho, \text{NR})} = \frac{\beta^{-1} + \tau}{1-p^- - p^+}.
\end{equation}
The above exact equation for the mean detection time admits a simple interpretation in the form of Bernoulli trials. We define each trial as an independent attempt at detection when the particle is initially non-reactive at one of the target's boundaries. Each attempt takes on average $\beta^{-1} + \tau$, where $\beta^{-1}$ is the mean time for the particle to turn reactive, and $\tau$ denotes the additional contribution coming from events where the particle is found outside the interval when it turns reactive, and thus has to return to its nearest boundary. The number of trials until detection follows a geometric distribution, and the mean number of trials is given by $(1-p_a-p_b)^{-1}$. This acts as a multiplicative factor to the mean time taken for one trial, and altogether yields the mean detection time.    

\subsection{Distribution of the detection time}

We now turn to compute the full distribution of the detection time. This is required to complement the limited information provided by the mean.

The calculation of the Laplace transforms of the distributions of the random variables in Eqs. (\ref{renewal_interval1}) and (\ref{renewal_interval2}) is done along the lines followed for their respective means (Sec. \ref{sub_mean_detection}). We have thus delegated the details of this calculation to Appendix \ref{Appendix:C}, and will hereby only quote the results. The Laplace transform of Eq. (\ref{renewal_interval1}) is

\begin{align} \label{distribution_1}
\tilde{T}_d(x_0,\sigma_{0},s)&=
\tilde{T}_f(\rho,s\mid x_0)\left[\pi_{\text{R}} + \pi_{\text{NR}}\tilde{T}_d(\rho,\textrm{NR},s)\right] 
 \\
&\pm (1-\pi_{\sigma_{0}})\tilde{T}_f(\rho,s+\lambda\mid x_0)\left[ \tilde{T}_d(\rho,\textrm{NR},s) -1 \right],\nonumber
\end{align}
where we have a plus sign if $\sigma_0=\textrm{NR}$, and a minus sign if $\sigma_0=\textrm{R}$. For $\sigma_0=\textrm{eq}$, the second term vanishes and we retain only the first term.

The Laplace transform of Eq. (\ref{renewal_interval2}) for the case $\rho=a$ is
\begin{equation} \label{distribution_2}
\tilde{T}_d(a,\text{NR},s)    = 
\frac{(\tilde{\phi}_a + \tilde{\chi}_a)(1- \tilde{\psi}^{+}_{b}) +  \tilde{\psi}^{+}_{a} (\tilde{\phi}_b + \tilde{\chi}_b)}{1 - \tilde{\psi}^{-}_{a} -\tilde{\psi}^{+}_{b} + \tilde{\psi}^{-}_{a}\tilde{\psi}^{+}_{b}  - \tilde{\psi}^{-}_{b}\tilde{\psi}^{+}_{a} } , 
\end{equation}
and for the case $\rho=b$ is
\begin{equation} \label{distribution_3}
 \tilde{T}_d(b,\text{NR},s)    = 
\frac{ (\tilde{\phi}_b + \tilde{\chi}_b)(1- \tilde{\psi}^{-}_{a})  + \tilde{\psi}^{-}_{b}(\tilde{\phi}_a + \tilde{\chi}_a)  }{1 - \tilde{\psi}^{-}_{a} -\tilde{\psi}^{+}_{b} + \tilde{\psi}^{-}_{a}\tilde{\psi}^{+}_{b}  - \tilde{\psi}^{-}_{b}\tilde{\psi}^{+}_{a} } ,
\end{equation}
where we have defined
\begin{equation} \label{defined_1}
 \tilde{\phi}_\rho(s) \equiv  \beta\int^{b}_{a}\tilde{C}(y,s+\beta\mid \rho) dy \equiv  \int^{b}_{a} \tilde{\Phi}_{\rho}(s+\beta) dy, 
\end{equation}
\begin{align} \label{defined_2}
&\tilde{\chi}_\rho(s) \equiv \int^{\infty}_{-\infty} \tilde{\Phi}_{\rho}(s+\beta) \times \\ &\Big[\pi_{\text{R}}\tilde{T}_f(\iota_{\pm},s \mid y) + \pi_{\text{NR}}\tilde{T}_f(\iota_{\pm},s+\lambda \mid y) \Big] 
 \Big[\Theta_{-}(y)+\Theta_{+}(y)\Big]dy ,  \nonumber
\end{align}
\noindent and
\begin{align} \label{defined_3}
\tilde{\psi}^{\pm}_{\rho}(s)  \equiv &  \int^{\infty}_{-\infty} \tilde{\Phi}_{\rho}(s+\beta) \times
\\
& \pi_{\text{NR}} \Big[ \tilde{T}_f(\iota_\pm,s \mid y) - \tilde{T}_f(\iota_\pm,s+\lambda \mid y) \Big]  \Theta_{\pm}(y)dy,   \nonumber
\end{align}
\noindent where again $\iota_- = a$ and $\iota_+ = b$.
As happened for the mean formulas, for the symmetric case in which the spatial dynamics and the boundary conditions to the left and the right of the target center are the same Eqs. (\ref{distribution_2}) and (\ref{distribution_3}) are equal and simplify considerably.  Setting $\phi_a = \phi_b := \phi$, $\chi_a = \chi_b := \chi$ and $\psi^{\pm}_{a} = \psi^{\pm}_{b}:=\psi^{\pm}$, we obtain
\begin{equation} \label{distribution_4}
  \Tilde{T}_d(\rho, \text{NR},s)   = \frac{\tilde{\phi} + \tilde{\chi}}{1 - \tilde{\psi}^{-} -\tilde{\psi}^{+} }.
\end{equation}


\subsection{Long time asymptotics -- Inheritance of power-laws} \label{subsec:long}

For simplicity, let us assume the symmetric case in which both the spatial dynamics and the boundary conditions to the left and the right of the target are the same. We focus on processes for which the first passage time distribution of the underlying ungated process has an asymptotic power-law behavior of the form 
\begin{equation} \label{heavytail}
    f(t) \simeq \frac{\theta}{\Gamma(1-\theta)} \frac{\tau_f^{\theta}}{t^{1+\theta}}, \quad 0<\theta<1,
\end{equation}
where $\tau_f>0$. Note that in this case, the mean first passage time diverges. By clever use of the Tauberien theorem, it can be shown that the small $s$ asymptotics of the Laplace transform of the first passage time is given by $\tilde{T}_f(\rho,s \mid y) \simeq1-(\tau_f s)^{\theta}$ (see pp. 43-45 in Ref. \cite{klafter2011first}). Note that $\tau_f$ is a function of the distance to the closest boundary of the interval target.

In Appendix \ref{Appendix:D}, we show that corresponding gated processes inherit the above asymptotics. That is to say, the power law $\theta$ remains
the same and the asymptotics differ only in the corresponding prefactor, which is determined exactly
\begin{equation} \label{asymptot1}
f_d(t,\rho, \text{NR}) \simeq
  \frac{\theta }{\Gamma(1-\theta)} \frac{\left(\pi^{-1}_{\text{NR}}\frac{B}{A}\right)}{t^{1+\theta}},
\end{equation}
where $A$ and $B$ are given in Appendix \ref{Appendix:D}. 

We thus see that, for the cases studied here, the gated detection time has a power law behaviour with the same $\theta$ of the corresponding ungated process, but with a different prefactor which can be determined exactly based on descriptors of the corresponding ungated process. In particular, for a single-point target ($b \to a$), we find
\begin{align} \label{asymptot2}
 f_d(t,a, \text{NR}) \simeq & \frac{1}{t^{1+\theta}} \frac{\theta}{\Gamma(1-\theta)} \times
 \\
 & \frac{\int^{\infty}_{-\infty} \tilde{\Phi}_{\rho}(\beta)\tau^{\theta}_f(y) dy}{\pi_{\text{R}} +\pi_{\text{NR}} \int^{\infty}_{-\infty} \tilde{\Phi}_{\rho}(\beta) \tilde{T}_f(a,\lambda\mid y) dy }, \nonumber 
\end{align}
where we recall that $\pi_{\textrm{R}}=\beta / \lambda$ and $\pi_{\textrm{NR}}=\alpha / \lambda$. In the converse limit of threshold crossing ($b \to \infty$) we obtain
\begin{align} \label{asymptot3}
 &f_d(t,a, \text{NR}) \simeq \frac{1}{t^{1+\theta}} \frac{\theta}{\Gamma(1-\theta)} \times
 \\
& \frac{\int^{a}_{-\infty} \tilde{\Phi}_{\rho}(\beta)\tau^{\theta}_f(y) dy}{\pi_{\text{R}} +\pi_{\text{NR}} \Big[\int^{\infty}_{a} \tilde{\Phi}_{\rho}(\beta)dy + \int^{a}_{-\infty} \tilde{\Phi}_{\rho}(\beta) \tilde{T}_f(a,\lambda \mid y) dy  \Big] } .  
\nonumber 
\end{align}

Note that the proof can easily be generalized to asymptotics of the form
\begin{equation} \label{heavytail2}
    f(t) \simeq \frac{\theta}{\Gamma(p-\theta)} \frac{\tau_f^{\theta}}{t^{1+\theta}} L(t), \quad 0<\theta<1,
\end{equation}
where $\tau_f>0$, and $L(t)$ is a slowly varying function, i.e., such that it fulfills $\lim_{t\to \infty} L(Ct)/L(t) = 1$ for all $C>0$ (for example, a constant function or a logarithmic function). The small $s$ asymptotics of the Laplace transform is then given by $\tilde{T}_f(\rho,s \mid y) \simeq 1 -(\tau_f s)^{\theta} L(1/s)$. 

\subsection{Transient behaviour under high crypticity}

In the work of Mercado-Vásquez and Boyer \cite{mercado2019first} a freely diffusing particle in search of a gated \textit{single-point target} was considered. The authors showed that when $K_{eq} = \frac{\alpha}{\beta} \gg 1$ an interim regime of slower power-law decay emerges before the asymptotic regime. A detection process for which $K_{eq} \gg 1$ was termed highly cryptic, since it spends most of its time in the non-reactive (hidden) state.

More recently, we have proved that under high-crypticity a slower transient regime is a universal feature of a wide range of gated processes in discrete space and time \cite{scher2020gated}. Yet, there remains a question: Is the condition $K_{eq} \gg 1$ sufficient to guarantee a slower transient regime for any gated process in continuous time, provided that the underlying ungated
process has the asymptotic power-law behavior of Eq. (\ref{heavytail})? In appendix \ref{Appendix:E}, we show that the answer to this question is no. Specifically, we show that whenever there is a possibility to spend some time on the target while being in the non-reactive state, an additional  condition is required to guarantee a slower pre-asymptotic power-law decay. 

Simply put, the time spent in the non-reactive state before transitioning to the reactive state must be considerably larger than the time spent on the target upon arrival. For the problem of a gated interval considered in this work, there is certainly a probability to spend time within the interval $[a,b]$ while being in the non-reactive state. In appendix \ref{Appendix:E}, we show that in this case the additional requirement translates to
\begin{equation} \label{crypric1}
 \frac{b-a}{\tau_r^\theta} \ll \beta^{\theta-1} ,
\end{equation}
with $\tau_r$ set by the small $s$ asymptotics of the spatial propagator $\tilde{C}(y,s\mid y) \simeq (\tau_r s)^{-\theta} $, and where we have assumed that $\tilde{C}(y,s\mid a) \simeq s^{-\theta} H(\abs{y-a}s^\theta)$, where $H$ is some function. Simple diffusion is just one example of a process that belongs to this group. In Appendix \ref{Appendix:E}, we also show that for such processes $\tau_f$ and $\tau_r$ are related by
\begin{equation} \label{taus_relation}
\tau_r = \left(\frac{\abs{y-a} \frac{d\tilde{C}(y,s\mid a)}{dy}}{\tau_f^{\theta}}\right)_{y=a} ^{-1/\theta}.
\end{equation}

Examining Eq. (\ref{crypric1}), it is thus clear that in the limit of single-point target, $b \to a$, the left-hand side of Eq. (\ref{crypric1}) is zero. The additional requirement is then fulfilled for every finite transition rate $\beta$. From the qualitative understanding above this is anticipated. Indeed, when the target is of measure zero, the particle spends no time on the target and can react with it only via crossing it while being the reactive state. Thus, in such cases we only care about the equilibrium occupancies of the reactive and non-reactive states. Namely, the rates can be arbitrarily large, as long as the rate to become reactive is much slower than its converse. On the other hand, Eq. (\ref{crypric1}) will never be satisfied in the case of threshold crossing, where $b \to \infty$. This means that the cryptic transient regime will never be observed in gated threshold crossing problems of the type analysed herein.

In appendix E, we show that when $K_{eq} \gg 1$ and the condition in Eq. (\ref{crypric1}) is satisfied the transient regime of the detection time density scales like
\begin{equation} \label{crypric2}
f_d(t,\rho, \text{NR}) \simeq \frac{1}{\Gamma(\theta)}\frac{A}{B}  t^{\theta-1}  ,  
\end{equation}
where $A$ and $B$ are the same as in the previous subsection (definitions can be found in Appendix \ref{Appendix:D}). Thus, under these conditions, we indeed see a transient regime with a different power law than the asymptotic. Furthermore, we can exactly determine this power-law, and its prefactor.

In appendix \ref{Appendix:E}, we also discuss the case of continuous time but discrete space, and specifically the model of gated CTRW on networks \cite{scher2020unifying}. We show that for such processes the additional requirement translates to $\gamma^{-1} \ll \beta^{-1}$, where $\gamma^{-1}$ is the average time spent on the target before leaving it, and $\beta^{-1}$ is the average time taken to transition from the non-reactive to the reactive state.

The findings presented in this subsection call for a refined definition of high-crypticity, which takes into account the time spent on the target: A process is highly-cryptic if it spends most of its time in the non-reactive state \textit{and} the time spent in the non-reactive state before transitioning to the reactive state is considerably larger than the time spent on the target upon arrival to it.


\section{Diffusion to a Gated Interval}\label{Sec:5}

In this section, we show how to apply the framework developed above to obtain explicit solutions for the detection time of a diffusing particle by a gated interval. We first solve for free diffusion and then consider the effects of confinement and drift.
\subsection{Freely diffusing particle}\label{sub:free}

Consider the gated interval problem illustrated in (Fig.~\ref{fig1}, bottom); and further assume that the particle, initially at $x_0$ at state $\sigma_{0} \in \{\text{R, NR}\}$, is not restricted and free to diffuse on the entire one-dimensional line. The gated target is the interval $[a,b]$.

The free-diffusion conserved propagator is a Gaussian and its Laplace transform is given by $ \tilde{C}(y, s | x_{0}) = \frac{1}{\sqrt{4 D s}} e^{-\sqrt{\frac{s}{D}} \abs{y-x_{0}}}$. Also, the Laplace transform of the first-passage time distribution from an arbitrary point $y$ to an arbitrary point $\rho$ is $\tilde{T}_f(\rho,s \mid y) = e^{-\sqrt{\frac{s}{D}}\abs{\rho-y}}$ \cite{redner2001guide}. In fact, these two classical results are all we need in order to solve the gated problem, using the renewal formalism established in Sec. \ref{Sec:4}. 

For any $x_0 \notin[a,b]$ we can break the trajectory of the gated problem into two parts. A first-passage trajectory to $\rho \in \{a,b\}$ that is followed, if the particle was not detected upon arrival, by a first detection process starting from the state $(\rho,\text{NR})$. Because the first part of the trajectory is a first-passage process which is well understood, from now on we will focus on the second part, i.e., a detection process with initial state $(\rho,\text{NR})$. It is important to note that accounting for the entire trajectory, by adding its first-passage part, is then a simple task that can generally be done via Eqs. (\ref{mean1}) and (\ref{distribution_1}).

It can be easily appreciated that in this example the spatial dynamics and the boundary conditions to the left and right of the target are the same, and we can use Eq.  (\ref{distribution_4}) to calculate the Laplace transform of the detection time distribution. Plugging $\tilde{C}(y, s | x_{0})$ and $\tilde{T}_f(\rho,s \mid y)$ into Eq. (\ref{distribution_4}) we obtain
\begin{equation} \label{diffusion1}
 \Tilde{T}_d(\rho, \text{NR},s) =   \frac{\frac{1}{\omega^{+}_{1}}-\frac{e^{-(b-a) \sqrt{\frac{s+\beta}{D}}}\left(\lambda+\omega^{-}_{1} \right)-\lambda}{\omega_2}}{2\beta^{-1} +\frac{\left(1+e^{-(b-a)} \sqrt{\frac{s+\beta}{D}}\right) \alpha \omega^{-}_{1}}{\omega_2 \left(\sqrt{\beta +s}+\sqrt{s}\right) \left(\sqrt{\beta +s}+\sqrt{\lambda +s}\right)}},     
\end{equation}
where we have defined $\omega^{\pm}_{1}=\sqrt{s+\beta}(\sqrt{s} \pm \sqrt{s+\lambda})$ and $\omega_2=(s+\beta) \lambda$. Similarly we can get the long-time asymptotics by plugging $\tilde{C}(y, s | x_{0})$ and $\tilde{T}_f(\rho,s \mid y)$ into  Eq. (\ref{asymptot1}). Noting that $\tau_f = \frac{\abs{y-x_0}^2}{D}$, we obtain
\begin{align} \label{free_diffusion2}
& f_d(t,\rho, \text{NR}) \simeq \frac{1}{2 \sqrt{\pi}} t^{-3/2} \times \\
  & \frac{\left(1+e^{-(b-a) \sqrt{\frac{\beta}{D}}}\right)(\sqrt{\beta}+\sqrt{\lambda}) \lambda}{2 \alpha \beta+\left(1-e^{-(b-a) \sqrt{\frac{\beta}{D}}}\right) \alpha \sqrt{\beta \lambda}+2\left(\beta^{2}+\sqrt{\beta^{3} \lambda}\right)}. \nonumber
\end{align}

Finally, in the previous section we have seen that for a finite-sized targets, high-crypticity requires the condition in Eq. (\ref{crypric1}) as well as $K_{eq} = \frac{\alpha}{\beta} \gg 1$. For our example here $\tau_r = 4 D$ and so the condition in Eq. (\ref{crypric1}), after squaring both sides of the equation, translates to
\begin{equation} \label{free_diffusion3}
 \frac{(b-a)^2}{4 D} \ll \beta^{-1} ,
\end{equation}
namely, the time spent in the non-reactive state before transitioning to the reactive state must be considerably larger than the time it takes the particle to diffuse a distance comparable to the size of the target. If both $K_{eq} \gg 1$ and Eq. (\ref{free_diffusion3}) hold, a transient regime emerges before the asymptotic regime of Eq. (\ref{free_diffusion2}). In this regime, according to Eq. (\ref{crypric2}), we have
\begin{align} \label{free_diffusion4}
 f_d(t,\rho, \text{NR}) \simeq & \frac{1}{\sqrt{\pi}} t^{-1/2} \times \\
&\sqrt{\beta}\left(\frac{2}{1+e^{-(b-a) \sqrt{\frac{\beta}{D}}}}-\frac{1}{1+\sqrt{K^{-1}_{eq}}}\right)   .  \nonumber
\end{align}

\begin{figure}[t!]
\centering
\includegraphics[width=\linewidth]{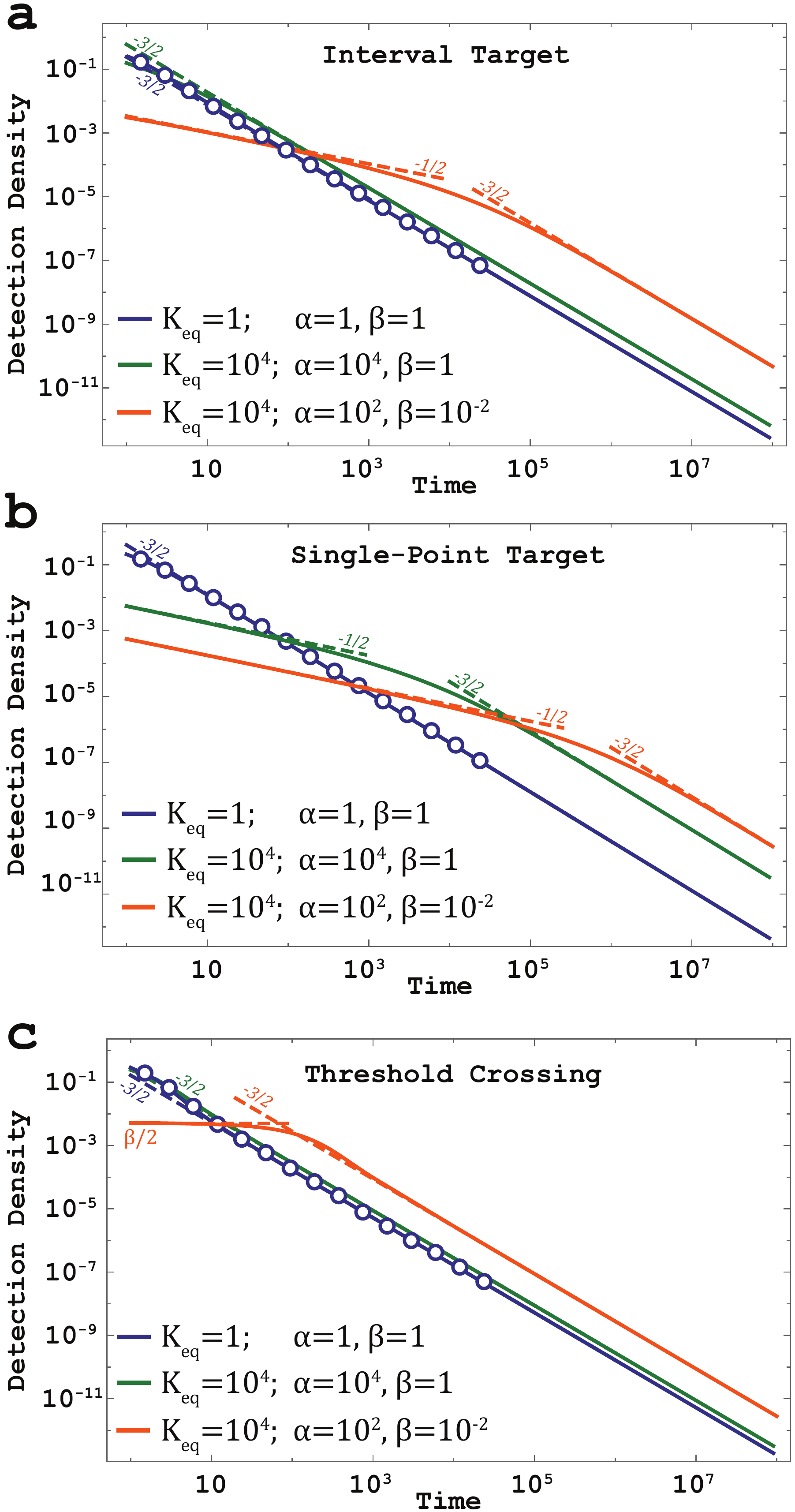}
\caption{A comparison of the detection time distribution and its dependence on the transition rates between models of (a) a gated interval of length $b-a=1$, (b) a gated point, and (c) gated threshold crossing. In all cases, we set $D=1$ for the diffusion coefficient, a non-reactive initial gate state, and $a$ for the initial position of the particle. In each panel, we plot three color coded curves, where each color represent a different choice of values for the transition rates $\alpha$ and $\beta$. For each choice of parameters, the lines represent numerical Laplace inversion of Eq. (\ref{diffusion1}), the dashed lines are the corresponding transient and asymptotic power laws according to Eqs.  (\ref{free_diffusion2}) and (\ref{free_diffusion4}), and circles come from Monte-Carlo simulations with $10^5$ particles and a simulation time step $\Delta t = 10^{-4}$.}
\label{fig2}
\end{figure}  

In Fig.~\ref{fig2} there are three panels corresponding to: 
(a) the gated interval problem, and its two extreme limits (b) the gated single-point target problem, and (c) the gated threshold crossing problem, which were illustrated in Fig.~\ref{fig1}. In each panel we plot three color coded curves, where each color represent a different choice of values for the transition rates $\alpha$ and $\beta$. Purple curves represent $\alpha=\beta=1$ such that $K_{eq}=\alpha/\beta=1$. Green curves represent $\alpha=10^4$ and $\beta=1$ such that $K_{eq}=10^4$. Orange curves represent $\alpha=10^2$ and $\beta=10^{-2}$ such that again $K_{eq}=10^4$. The green and orange curves have the same ratio of transition rates ($K_{eq}=10^4$), but for the latter transitions are much slower, and specifically $\beta$ is much smaller. 

In panel (a), the case of an interval target, we see that the purple and green curves are similar in shape --- an asymptotic $\sim t^{-3/2}$ power-law kicks in rather early. In contrast, the orange curve possesses a prolonged transient regime of $\sim t^{-1/2}$ before the asymptotic regime enters. This is because the orange curve fulfills both conditions for high-crypticity, i.e., $K_{eq} \gg 1$ \textit{and} Eq. (\ref{free_diffusion3}). 

In panel (b), the case of a single-point target, the green curve is actually similar in shape to the orange curve. In this limit the condition in Eq. (\ref{free_diffusion3}) is always met and it is sufficient to require that $K_{eq} \gg 1$. Note however, that while the ratio of the transition rates alone determines whether a prolonged transient regime exists, the duration of this regime is affected by the magnitude of the rates through $\lambda=\alpha+\beta$. More specifically, the transition between the transient and asymptotic regimes occurs at $K_{eq}^2/\lambda$ \cite{mercado2019first}.

Finally, in panel (c) we present the case of threshold crossing. There, we see that non of the curves posses a prolonged transient regime, as the condition in Eq. (\ref{free_diffusion3}) is never met. Furthermore, for short times (up to $\beta^{-1}$) the detection probability density is constant with a value of $\beta/2$. This can be easily understood by the following argument: With probability $\beta exp (- \beta t) \simeq \beta$ the particle becomes reactive after a short time $t$. Since the motion is symmetric, upon becoming reactive the particle has a probability half to be found above its starting position. Recalling the particle started on the threshold, it has a probability half of being detected upon becoming reactive. In total, the first detection probability density at short times is $\beta/2$.

\subsection{Diffusion in confinement}\label{sub:confinement}

Let us now restrict the particle to diffuse inside a box $[0,L]$ with reflecting boundaries, such that the gated target is inside the box ($0<a \leq b<L$). The confinement renders the first-passage asymptotics exponential, so the results derived in Sec. \ref{Sec:4} for heavy-tailed distributions are no longer valid. However, it also renders the mean first-passage time finite, and so one can use Eq. (\ref{mean1}) together with Eqs. (\ref{mean9}) and (\ref{mean10}) to obtain the mean detection time.

Furthermore, as we did for the freely diffusing particle, we can of course calculate the entire Laplace transform of the reaction time. Let us again focus on a reaction with initial state $(\rho,\text{NR})$, i.e., assume the particle starts on the boundary in the non-reactive state. For simplicity, let us further assume that the center of the gated interval is situated exactly at the center of the confining box: $a=(L-l)/2$ and $b=(L+l)/2$, such that $l=b-a$. We can thus use Eq. (\ref{distribution_4}). All we require for this calculation is the conserved spatial propagator for diffusion restricted to a box $[0,L]$, and the first passage time to $a$, starting from some $x \in [0,a]$ (which in our case is equal to the first passage time to $b$ starting from $L-x$). These quantities are calculated in Appendix \ref{Appendix:F}.

In Fig.~\ref{fig3}, we set $\alpha=1$, $L=10$, $l=1$ and $D=1$, and plot the detection time probability density for $\beta=1$ (blue line, $K_{eq}=1$) and for $\beta=10^{-2}$ (orange line, $K_{eq}=10^{2}$). For each case, we also plot an exponential distribution with the mean taken to be the mean detection time calculated according to Eq. (\ref{mean11}) (dashed lines of corresponding colors).  For $\beta=1$ the distribution is clearly non-exponential, there are two distinct phases. Therefore, despite having an exponential tail, the gated distribution cannot be captured by a single exponent, which is to be expected since multiple time scales are involved in the problem. However, for $\beta=10^{-2}$ high-crypticity conditions are met, and we observe Poisson-like asymptotics \cite{godec2016universal}. This is to say that the distribution is well approximated by an exponential distribution whose mean is simply the mean detection time. This can be understood by noting that the time it takes for the diffusing particle's position to equilibrate over the box is much shorter than the time it takes the particle to turn reactive. The latter then becomes the rate-limiting step which dominates the distribution of detection times. This example warrants caution --- Poisson kinetics is not guaranteed in the general case, but does emerge in the cryptic regime. \\ 

\begin{figure}[t]
\centering
\includegraphics[width=\linewidth]{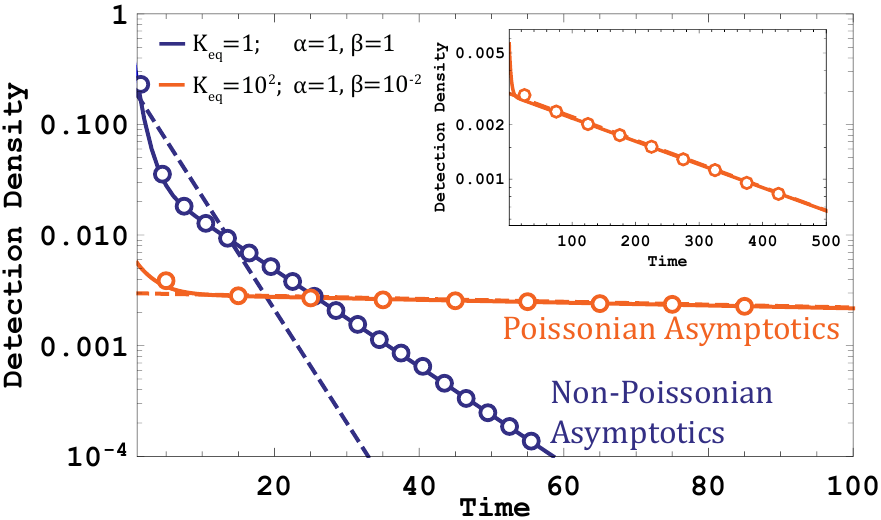}
    \caption{The detection time distribution at a gated interval $[a,b]$, for a diffusing particle restricted to a box $[0,L]$ with reflecting boundaries. The lines are numerical Laplace inversions of Eq. (\ref{distribution_4}), where $\phi$, $\chi$ and $\psi^{\pm}$ are calculated using the results of Appendix \ref{Appendix:F}. The circles are the results of Monte-Carlo simulations with $10^5$ particles and a simulation time step $\Delta t = 10^{-4}$. Here, we take: $\alpha=1$, $L=10$, $D=1$ and $l=1$ where $a=(L-l)/2$ and $b=(L+l)/2$. The gate is initially in the non-reactive state, and we set $a$ for the initial position of the particle. The blue line is drawn for the case $\beta=1$ and $K_{eq}=1$, and the orange line is drawn for the case $\beta=10^{-2}$ and $K_{eq}=10^{2}$.  The dashed lines represent exponential distributions whose means are taken to be the mean detection times according to Eq. (\ref{mean11}). While a single exponential is not expected a priori, it is observed for the case of high-crypticity (orange). It can be appreciated that the distribution is well described by the dashed line (also see inset).}
    \label{fig3}
\end{figure}


\subsection{Diffusion with drift} \label{Sec:6}

As we discussed above, the mean first-passage time of a freely diffusing particle diverges, and this property is inherited by the corresponding gated problem. Confinement can regularize the mean and make it finite. This can also be done by introducing a constant drift velocity $v$ in the target's direction. The mean first-passage time of the ungated problem is then simply given by $\ell/v$, where $\ell$ denotes the distance between the initial position of the particle and the point target. However, considering the gated counter-part of this problem, we observe that the mean detection time diverges despite the constant drift. This fact can be intuitively understood through the following argument: when the drift drives the particle downhill towards the target, there is a non-zero probability that the particle will arrive at the target in the non-reactive state. Subsequently, by the time it turns reactive again the particle is likely to be on the other side of the target, and it now has to travel ``uphill" for the detection to occur. This renders the mean detection time infinite.  It is thus clear that the mean detection time under stochastic gating can be remarkably different from its ungated counterpart.

\begin{figure}
    \centering
    \includegraphics[width=\columnwidth]{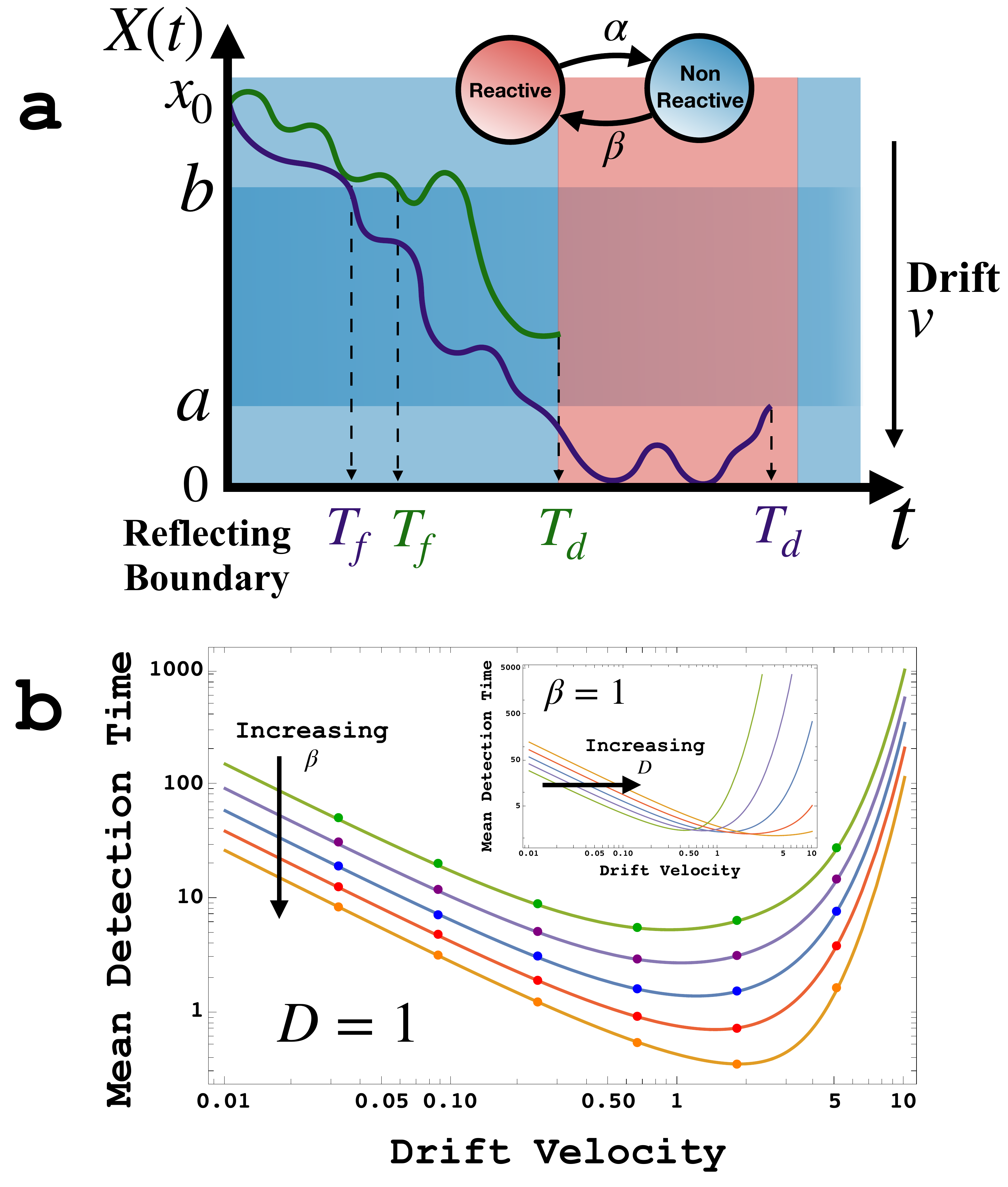}
    \caption{Detecting a particle diffusing with drift by a gated interval. (a) Schematics of the process where a particle, initially at $x_0$ while the gate is in the non-reactive state, is diffusing on the positive ray $(0,\infty)$ with a drift velocity $v$ towards the origin. The gated interval is $[a,b]$ and the origin is considered reflective. The two trajectories represent two different types of detection events. The purple trajectory illustrates a scenario likely to happen when $v$ is large. Namely, if the particle arrives at the upper boundary $b$ when the gated interval is non-reactive, then the particle may be able to cross the interval without being detected. It will subsequently need to go against the drift for a detection event to occur. The green trajectory is representative of low $v$, where a particle that arrives at the upper boundary when the interval is non-reactive is unlikely to cross the interval without being detected. (b) Mean detection time $\langle T(b,NR) \rangle $ vs. $v$ for $\beta = 0.25, 0.5, 1, 2,$ and $4$, at $\alpha=D=1$ . The mean detection time displays a non-monotonic dependence on $v$, and achieves a minimum for some $v=v^*$. Inset shows plots for $D=0.25, 0.5, 1, 2,$ and $4$, at $\alpha=\beta=1$, showing that $v^*$ also depends on $D$.}
    \label{fig:schematic_bounded}
\end{figure}

A slight variation of the drift-diffusion model discussed above can nevertheless render the gated mean finite. In particular, consider a particle diffusing under a constant drift velocity $v$ towards the origin, with the origin also being a reflective boundary. For $0<a<b$, we consider a gated interval $[a,b]$, where the particle can get detected and absorbed in its reactive state. Despite the constant drift, the reflecting boundary at the origin ensures that the mean first-passage time to the boundaries of the interval remains finite, irrespective of whether the particle has to travel uphill or downhill. Consequently, the mean detection time is also finite. A schematic of this setup is provided in Fig.~\ref{fig:schematic_bounded}a.

The formalism developed in Sec. \ref{Sec:4} asserts that knowing certain ungated observables is enough in order to obtain the first detection time statistics. In particular, a key quantity is the conserved propagator for the diffusion equation with drift. This obeys
\begin{equation} \label{drift}
    \frac{\partial C(x, t  | x_{0} )}{\partial t}=D \frac{\partial^{2}C(x, t | x_{0})}{\partial^2 x} + v \frac{\partial  C(x, t | x_{0})}{\partial x},
\end{equation}
with the initial condition $C(x, t=0 | x_{0})=\delta\left(x-x_{0}\right)$ and boundary conditions $\frac{dC(x, t | x_{0})}{dr} \big|_{x=0} = 0$ and $C(x \to \infty, t  | x_{0} ) = 0$. The drift is $v>0$ and its direction is towards the reflecting boundary at zero. In appendix \ref{Appendix:H}, we obtain $ \tilde{C}(x, s | x_{0})$ in Eq. (\ref{H:2}).  The Laplace transform of the first-passage probability $\tilde{T}_f(x, s | x_{0})$ and its mean $\braket{T_f (x| x_{0})}$ can also be obtained and we give them in Eqs. (\ref{H:11}) and (\ref{H:12}). Assuming for simplicity that the particle starts at the boundary $b$, one can utilize Eq. (\ref{mean10}) to obtain the mean detection time $\langle T_d(b,NR) \rangle$ by plugging in the ungated quantities obtained above.

For intermediate values of $v>0$, it is clear that drift can speed up detection as it helps avoid situations where the particle drifts away from the interval. However, if $v$ is sufficiently large, a significant contribution to the detection time comes from trajectories where the diffusing particle crosses over to the other side of the interval (with its position somewhere between $0$ and $a$), and then travels uphill, against the drift, for the eventual detection (see Fig. \ref{fig:schematic_bounded}a and the associated caption). Thus, one would expect the mean detection time $\langle T_d(b,NR) \rangle$ to vary non-monotonically as a function of $v$. This expectation is indeed verified in Fig.~\ref{fig:schematic_bounded}(b), where we fix $\alpha=D=1$, $a=1$ and $b=3$ and plot $\langle T_d(b,NR) \rangle$ vs. $v$ for $\beta=0.25, 0.5, 1, 2,4$. For small values of $v$,  the mean detection time decreases linearly as the drift is increased. However, for large $v$, it increases rapidly indicating that detection is much more difficult. 

This naturally leads to the question of finding the optimal drift velocity which minimizes the mean detection time. From Fig.~\ref{fig:schematic_bounded}(b), it is evident that the value of $v^*$, which is the value of $v$ for which $\langle T_d(b,NR) \rangle$ achieves its minimum value, increases as $\beta$ increases. This is expected as increasing $\beta$ increases the amount of time the particle spends in the reactive state. This, in turn, reduces the chance that the particle will cross the interval undetected, which allows for higher drift velocities. Naively, one could formulate the following argument to find $v^*$: the mean time taken for the particle to turn reactive is $\beta^{-1}$, while the particle travels an average distance of $v/\beta$ during this time. So, one could give a preliminary estimate of $v^* \approx (b-a)\beta$. However, as we show in the inset of Fig.~\ref{fig:schematic_bounded}(b), $v^*$ also depends on the diffusion coefficient, with smaller values of $D$ corresponding to higher values of $v^*$. This highlights the importance of the exact result obtained in Eq. (\ref{mean10}) which captures the explicit dependence of the mean detection time on $D$, along with other relevant parameters, and allows us to analytically study this optimization problem. \\




\section{Diffusion to a Gated Sphere}\label{Sec:7}

In Sec. \ref{Sec:4}, we developed a general framework to treat one-dimensional gated processes. We have not been able to generalize this theory to cases of arbitrarily shaped targets in higher dimensions. However, if the target and spatial dynamics are rotationally symmetric, such a generalization is possible: Instead of a gated interval $[a,b]$ one can consider a gated annulus, or spherical shell, with some inner radius $r_a$ and outer radius $r_b$. Furthermore, one can study a gated disk, or sphere, by taking the limit $r_a \to 0$ (The corresponding sphere in one-dimension would be the interval $[0,b]$, where there is a reflecting wall at the origin). In this case, the equations for the mean and distribution of the detection time simplify considerably, since exiting the target through $r_a$ is impossible, and so all contributions associated with such trajectories vanish.

Moreover, the case of d-dimensional free diffusion and a spherically symmetric target, can be further simplified by utilizing the well-established mapping between the distance from the origin of d-dimensional free diffusion (Bessel process) and one-dimensional diffusion in a logarithmic potential \cite{bray2000random,martin2011first,ryabov2015brownian,ray2020diffusion}. The gated first detection time for the latter can then be is easily attained with the formalism of Sec. \ref{Sec:4}, and the solution can be mapped back onto the d-dimensional case.

Next, we take time to explain the mapping by writing the corresponding Fokker-Planck equations and comparing them. Starting with the equation for a freely diffusing particle in d-dimensions, we note that the propagator does not depend on the angular part of the Laplacian. Hence, we are left only with the radial part $\frac{1}{r^{d-1}}\frac{\partial}{\partial r} \left( r^{d-1} \frac{\partial}{\partial r}\right)$, where the distance from the origin is denoted by $r \equiv \abs{\vec{r}}$. The d-dimensional diffusion equation can thus be written as
\begin{equation} \label{radial_diffusion}
    \frac{\partial  C(\vec{r}, t \mid  \vec{r}_0)}{\partial t}  =   D\frac{d-1}{r} \frac{\partial C(\vec{r}, t \mid \vec{r}_0)}{\partial r} +D \frac{\partial^{2} C(\vec{r}, t \mid \vec{r}_0)}{\partial r^{2}}. 
\end{equation}
Equation (\ref{radial_diffusion}) is written for the propagator. Here we are interested in $C(r, t \mid  r_0) =\Omega_d r^{d-1} C(\vec{r}, t \mid  \vec{r}_0) $, where  $\Omega_d$ is the surface area of a d-dimensional \textit{unit} sphere \cite{redner2001guide}. By plugging this relation into  Eq. (\ref{radial_diffusion}) we obtain the Fokker-Planck equation for $C(r, t \mid  r_0)$:
\begin{align} \label{radial_diffusion_dist}
 \frac{\partial  C(r, t \mid  r_0)}{\partial t}  =   \frac{\partial}{\partial r}\Big[\left(D\frac{1-d}{  r}\right)& C(r, t \mid r_0)\Big]
    \\
    &+D \frac{\partial^{2} C(r, t \mid r_0)}{\partial r^{2}}. \nonumber
\end{align}

We will now show that Eq. (\ref{radial_diffusion_dist}) is similar in form to the Fokker-Planck equation for a particle diffusing on the semi-infinite line in a logarithmic potential $U(x) = U_0 \text{log} \abs{x}$, where $U_0$ has units of energy and $x$ is dimensionless. At $x=0$ the wall is completely reflective. Given that the particle has started at $x_0$, the conserved spatial propagator follows the Smoluchowski diffusion equation
\begin{align} \label{log1}
    \frac{\partial  C(x, t \mid  x_0)}{\partial t}  =   \frac{\partial}{\partial x}\Big[\left(\frac{U_{0}}{\zeta x}\right)& C(x, t \mid x_0)\Big]
    \\
    &+D \frac{\partial^{2} C(x, t \mid x_0)}{\partial x^{2}}, \nonumber
\end{align}
with initial condition $C(x,0 \mid x_0) = \delta(x-x_0)$ and a boundary conditions $\left[D\frac{\partial C(x, t | x_{0})}{\partial x} + \frac{U_0}{\zeta x}C(x, t | x_{0})\right]_{x=0} = 0$ and $C(x \to \infty, t  | x_{0} ) = 0$, where $\zeta$ is the friction coefficient such that $ D = k_B T / \zeta$.

\begin{figure}[t!] 
\centering 
\includegraphics[width=\linewidth]{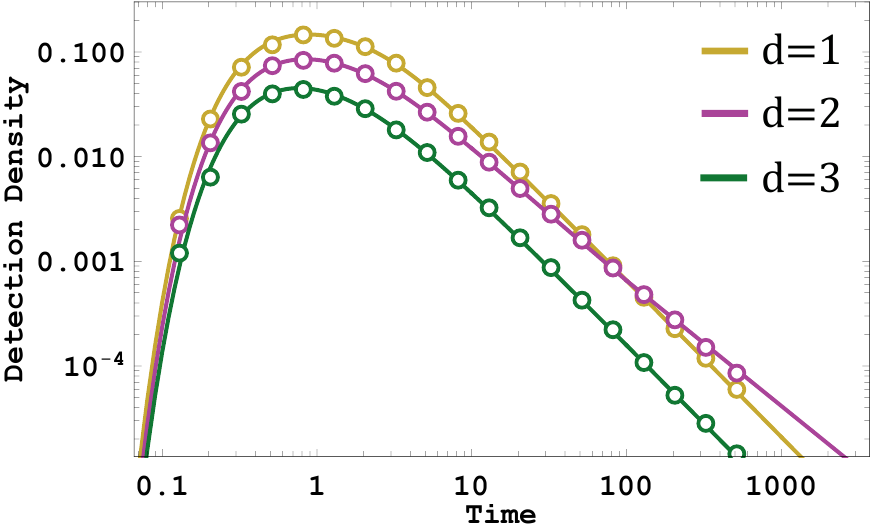}
\caption{The detection time density where the target is a gated d-dimensional unit sphere centered at the origin. The particle is freely diffusing, and we have set $\alpha=\beta=1$, $D=1$, $r_0=3$ and $\sigma_0=\textrm{eq}$. We consider diffusion in $d \in \{1,2,3\}$ dimensions. For each $d$, the lines are numerical Laplace inversions of Eq. (\ref{distribution_1}) to which we have plugged in Eq. (\ref{log2}). Circles come from Monte-Carlo simulations with $10^5$ particles and a simulation time step $\Delta t = 10^{-4}$. }\label{fig4}
\end{figure}  

Indeed, Eq. (\ref{log1}) can be mapped onto Eq. (\ref{radial_diffusion_dist}) by taking $x \to r$  and  $\frac{U_{0}}{D\zeta} \to 1-d$. In the following we will use the formalism of Sec. \ref{Sec:4} to obtain the detection probability for the gated version of the one-dimensional logarithmic potential problem, and then use the aforementioned mapping to obtain the detection time distribution of diffusion in d-dimensions by a gated sphere centered at the origin.

Assuming an interval target $[0,b]$, the detection time of a diffusing particle on the semi-infinite line in logarithmic potential, where $x_0 = b$ and $\sigma_0 = \text{NR}$, is given by Eq. (\ref{distribution_3}). In fact, since exiting the target through $a=0$ is impossible (there is a reflecting wall at $x=0$), Eq. (\ref{distribution_3}) simplifies substantially
\begin{equation} \label{log2}
 \Tilde{T}_d(b,\text{NR},s) = \frac{ \tilde{\phi}_b + \tilde{\chi}_b  }{1 -\tilde{\psi}^{+}_{b}},
\end{equation}
where $\phi_b$, $\chi_b$ and $\psi_b^{+}$ were defined in Eqs. (\ref{defined_1}), (\ref{defined_2}) and (\ref{defined_3}) respectively. These functions require in turn the conserved spatial propagator and the first-passage time to a point $b$, which we calculate in Appendix \ref{Appendix:G}. Plugging these results into  Eq. (\ref{log2}) gives the gated detection time given the initial condition $(b,\text{NR})$. To obtain the detection time for a general initial condition, we plug this result into Eq. (\ref{distribution_1}). Finally, by identifying $x$ with $r$ and  $d$ with $1-\frac{U_{0}}{D\zeta}$ we obtain the detection time distribution for diffusion in d-dimensions by a gated sphere of radius $r_b$ centered at the origin. In Fig.~\ref{fig4}, we plot the detection time of a freely diffusing particle by a d-dimensional unit sphere, for $d=1,2,3$, with $\sigma_0=\textrm{eq}$ and $r_0=3$.

\begin{figure}[t!] 
\centering 
\includegraphics[width=\linewidth]{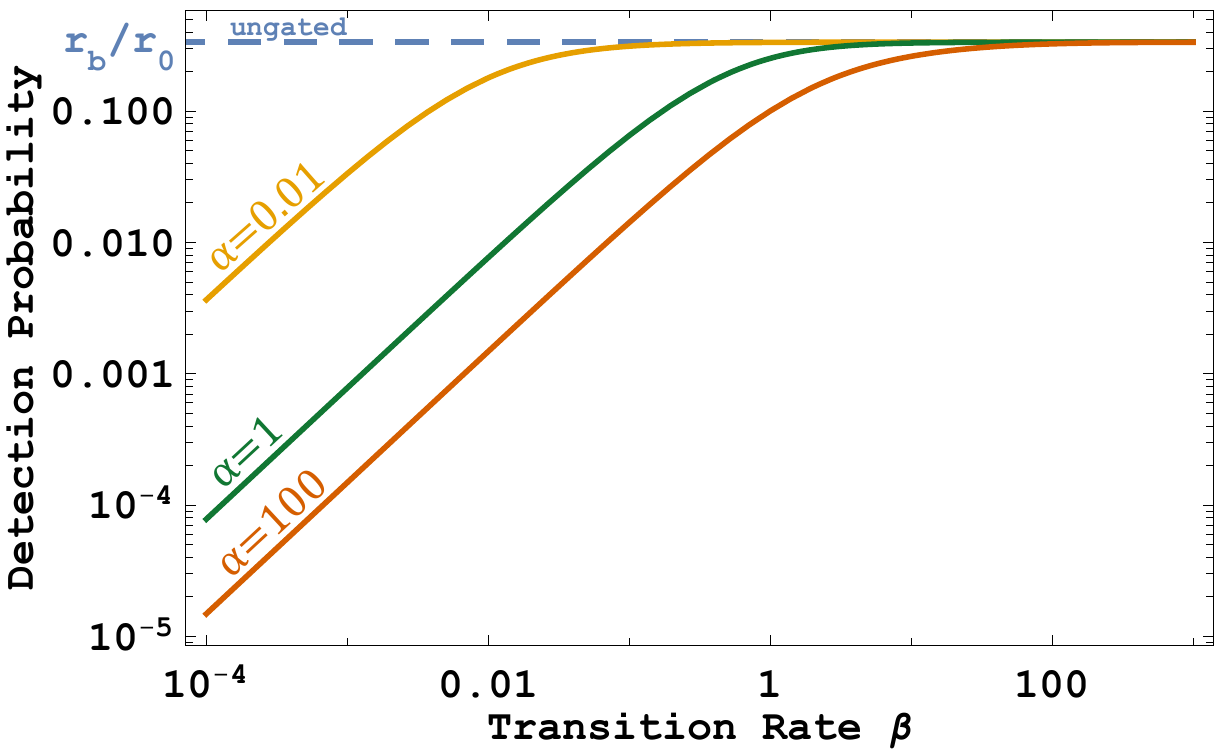}
\caption{Detection probability by a three-dimensional unit sphere. We plot the detection probability, namely the probability to be detected eventually, vs. the transition rate $\beta$. Each  curve represents a different value of the transition rate $\alpha$. The dashed horizontal line is the ungated benchmark, where the detection probability is $r_b/r_0$. Here, we used $D=1$, $\sigma_0=\textrm{eq}$ and $r_0=3$.}\label{fig6}
\end{figure} 

 The one and two-dimensional ungated first-passage processes are recurrent, the probability to eventually arrive at the sphere is one. Thais can be seen by looking at the small $s$ expansion of the Laplace transform of the first-passage distribution, and noting that the first term in this expansion is $1$. By definition, this term corresponds to $\int_{0}^{\infty} f(t \mid \vec{x})dt$, where $f$ is the first-passage distribution. However, the three-dimensional process is transient, the probability to eventually arrive at the sphere is $r_b/r_0$ (the ratio between the spherical target radius and the initial distance from the origin), which can be strictly smaller than one.
 
Recall that the asymptotic form of the one-dimensional first-passage distribution fits exactly the form assumed in Sec. \ref{subsec:long} for the inheritance of the power-law. The proof can be easily generalized to include the three-dimensional case, and with more care to account for the logarithmic corrections of the two-dimensional case as well.  The detection probability by the d-dimensional sphere can also be obtained for the Laplace transform of the detection time. As expected, in one and two dimensions we find that the particle will eventually be detected with probability one (regardless of the gating rates). However, intriguingly, while in the ungated three-dimensional case the detection probability is only dependent on the ratio $r_b/r_0$, in the corresponding gated case the detection probability is also a function of the transition rates and the diffusion coefficient. In particular, taking $\beta \to 0$, or $\alpha \to \infty$, we find that the detection probability vanishes. This finding is illustrated in Fig. (\ref{fig6}).

\section{Conclusions} \label{Sec:8}

\subsection{Summary}

Gated first-passage processes, which arise in various situations ranging from chemical reactions to the analysis of time-series data, were the focal point of this paper. In Sec.~\ref{Sec:2}, we surveyed the state of the art: the SSLW approach to continuous gated first-passage processes and showed how it can be used to compute detection time statistics for gated point targets. We then outlined, in Sec.~\ref{Sec:3}, why this approach fails to generalize to more realistic scenarios, and in particular why it leads to an apparent impasse when considering gated targets of non-zero length, area, or volume. In Sec.~\ref{Sec:4}, we presented a novel renewal framework that circumvents obstacles that arise in the SSLW approach, and subsequently yields closed-form solutions for the statistics of the detection time density in terms of the
ungated propagator and first-passage time. In particular, this renewal approach allowed us to obtain the Laplace transform of the detection time density and all its moments. 

Crucially, we showed that the exact results derived herein shed light on universal features of gated first-passage processes. Namely, in situations where the ungated first-passage time density is characterized by power-law asymptotics, the corresponding detection time density inherits the same power-law decay, albeit with a different prefactor. The long-time power-law tail may be preceded by a slower transient power-law decay with a different exponent. Our formalism reveals that, in the case of point targets, such a transient power-law decay is a generic feature of Markovian gated first-passage processes. Yet, an additional condition is required to guarantee its existence for targets of non-zero volume.

In Sec.~\ref{Sec:5}, the power of our framework was demonstrated in several examples. Specifically, we provided the first analytical solutions to the detection time problem of a diffusing particle by a gated interval. These solutions cover the cases of free and confined diffusion, as well as diffusion with drift. Importantly, going beyond point targets revealed an interplay between the time it takes the particle to become reactive and the time it takes it to diffuse across the target, which leads to a refined definition of the cryptic pre-asymptotics regime. Confinement and drift also play a role, but while the former always leads to exponential detection time tails, the latter may or may not. Crucially, and in contrast to the ungated case, a particle that is drifting towards a gated interval may cross it undetected. Boundary conditions downstream will then determine whether the particle is eventually detected, or has a chance to escape forever. We show that a reflecting boundary downstream is enough to guarantee detection, and that in this case there is an optimal drift velocity that minimizes the mean detection time.  

Finally, in Sec.~\ref{Sec:7} we demonstrated the broader scope of our framework by obtaining the first analytical solutions for the detection time statistics by a gated disk or sphere. This was done by leveraging the mapping between d-dimensional simple diffusion and one-dimensional diffusion in the presence of a logarithmic potential, the latter falling directly within the realm of applicability of our framework. We stress that in this case and others, using our approach to obtain the desired solution to a gated problem, only requires plugging in the propagator and first-passage time of the ungated problem. Thus, when a solution to the ungated problem is known analytically, or can be estimated numerically, a solution to the corresponding gated problem is now also readily available.    

\subsection{Outlook}

The renewal framework introduced in this paper, and the consequent results derived from it open a wide range of theoretical and experimental directions for further inquiry. 

This paper is focused on continuous processes. Yet, in fact, our formalism can also treat processes with discontinuous trajectories (albeit still in continuous space). An interesting example is stochastic jump processes \cite{gardiner1985handbook}, which can be treated by our formalism as long as jumps cannot be made into the gated target itself. Consider for example continuous stochastic processes that undergo stochastic resetting. We note that for the gated point target search and threshold crossing problems, the formulas derived in this paper hold without change, even in the presence of resetting \cite{evans2011diffusion,evans2020stochastic,renewal2,renewal3,renewal4,renewal5,renewal6,renewal7,renewal9,renewal11,renewal13,kumar2023universal,renewal15,renewal16,renewal17,bonomo2021first,pal2017first,pal2019first,bressloff2020modeling,ray2020space,evans2020stochastic,pal2022inspection,renewal10}. Generalization to the case of a gated interval is also straightforward, but requires careful consideration since restart can teleport the particle from one side of the interval to another without need to cross the interval itself.

The fascinating interplay between the effects of resetting and gating, forms a fertile ground for stimulating research in the domain of gated chemical reactions, as well as the extremal and record statistics of partially observable time-series which undergo restart. From a theoretical standpoint, several exciting curiosities emerge. The literature on resetting and gating benefited immensely from unified renewal approaches that were introduced to facilitate the analysis of such processes. However, when combined, the order in which these renewal frameworks are applied changes the problem description and its dynamics. For example, and as we have verified (results not shown), by first computing the statistics of diffusion on the semi-infinite line under resetting, and then plugging the results into the gating renewal formalism developed herein, one gets the setting described and analyzed in Refs.~\cite{mercado2021search} and ~\cite{biswas2023rate}, where the internal dynamics of the system are decoupled from the spatial processes under restart. However, applying the renewal formalisms in the opposite order, i.e., obtaining the gated dynamics first and then plugging it into the restart renewal framework leads to the setting considered in Ref.~\cite{bressloff2020diffusive}, where upon restart, the composite process, consisting of the underlying process as well as the gating dynamics, are reset to a predefined state. 

Before concluding, we stress that the theory developed herein is a single-particle one, namely we consider a single particle in search of a single target (also known as the isolated pair problem). We caution that one should not naïvely take our results and apply them to the multi-particle (single target still) case. This is especially tempting, since Smoluchowski's original approach to diffusion-limited reactions was to describe the many-body problem as an effective pair problem. This approach essentially relies on the assumption that the kinetics of the single particle is not influenced by other surrounding particles, i.e., that there are no correlations whatsoever between the different particles. However, in 1996, Zhou and Szabo shared their realization that this is only true to the case of gated particles, whereas gating of the target renders the particles correlated (unless the gating is very fast or the gate dynamics is ``frozen", i.e., in the limit $\alpha,\beta \to 0$, when only one state is initially occupied) \cite{berezhkovskii1997smoluchowski,zhou1996theory}. Berezhkovskii, Yang, Sheu and Lin extended the result to the case of many gate states and demonstrated (by elegantly invoking Jensen's inequality) that the reaction is faster when the gating is of the particles \cite{berezhkovskii1997smoluchowski, berezhkovskii1996stochastic}. Recent works by Bressloff and Lawley showed similar phenomenon for particles reacting with a gated boundary \cite{bressloff2015stochastically,bressloff2015moment}. They also considered the complementary problem of many gated targets (patches on a surface) and a single particle \cite{bressloff2015escape2}. 

Finally, an important generalization of the framework considered herein is to the case of non-Markovian gating. Indeed, it is often the case that the dynamics of molecular gates are governed by binding and unbinding events of ligands. In turn, these events themselves are governed by first-passage processes whose first-passage times distributions need not be exponential. This highlights the importance of understanding the role non-Markovianity plays in determining the statistics of gated detection times. Non-Markovian gating also plays a central role when considering periodically sampled time-series, which have been recently shown to have interesting and distinct behaviors compared to their ungated counterparts. Developing an analytical framework to address these questions is an important direction for future research. 

\emph{Acknowledgments.---}A.K. gratefully acknowledges the Prime Minister’s Research Fellowship of the Government of India for financial support. M.S.S. acknowledges the support of a MATRICS Grant from SERB, Government of India. This project has received funding from the European Research Council (ERC) under the European Union’s Horizon 2020 research and innovation program (Grant agreement No. 947731).\\

\noindent Y.S. and A.K. contributed equally to this work. 
\appendix
\section{Propagator for simple diffusion on the positive half line with a reflecting boundary condition at the origin} \label{Appendix:A} \setcounter{equation}{0}

For completeness, we compute the propagator for the diffusion equation with the initial condition $C(x, t=0 | x_{0})=\delta\left(x-x_{0}\right)$ with $x_{0}>0$, and boundary conditions $\frac{dC(x, t | x_{0})}{dx} \big|_{x=0} = 0$ and $C(x \to \infty, t  | x_{0} ) = 0$.

Laplace transforming the diffusion equation we have
\begin{equation} \label{A:1}
    s \tilde{C}(x, s  | x_{0} )-\delta\left(x-x_{0}\right)=D \frac{d^{2}\tilde{C}(x, s | x_{0})}{dx^2}.
\end{equation}

\noindent In each subdomain, $x < x_{0}$ and $x > x_{0}$, the solution is a linear combination of the exponential functions $\text{exp}(\sqrt{\frac{s}{D}}x)$ and $\text{exp}(-\sqrt{\frac{s}{D}}x)$. Due to the boundary condition at $x = 0$, the linear combination coefficients are equal for $x < x_{0}$, and so we can write $\tilde{C}(x, s  | x_{0})=A\hspace{1pt}\text{cosh}(\sqrt{\frac{s}{D}}x)$. For  $x > x_{0}$ the boundary condition at $x \to \infty $ asserts that only the decaying exponential is possible. Altogether we have
\begin{equation} \label{A:2}
 \tilde{C}(x, s | x_{0}) = \begin{cases}
 \tilde{C}_{<} = A(s)\hspace{1pt}\text{cosh}(\sqrt{\frac{s}{D}}x),  \hspace{17pt} x < x_{0},\\
 \tilde{C}_{>} =B(s)e^{-\sqrt{\frac{s}{D}}x}, \hspace{37pt}  x > x_{0}.\\
 \end{cases}    
\end{equation}

Imposing continuity of the concentration at $x = x_{0}$ and also the joining condition $
     -1=D\Big[\frac{\partial \tilde{C}_{>}}{\partial x}\big|_{x=x_{0}}  - \frac{\partial \tilde{C}_{<}}{\partial x}\big|_{x=x_{0}}\Big]$, we obtain
\begin{equation} \label{A:3}
 A(s)=\frac{1}{\sqrt{s D}} e^{-\sqrt{\frac{s}{D}}x_{0}},
\end{equation}
and 
\begin{equation} \label{A:4}
 B(s)=\frac{1}{\sqrt{s D}} \text{cosh}\Bigg(\sqrt{\frac{s}{D}}x_{0}\Bigg).
\end{equation}

Plugging Eqs. (\ref{A:3}) and (\ref{A:4}) back into Eq. (\ref{A:2}) we obtain the propagator in Laplace space
\begin{equation} \label{A:5}
 \tilde{C}(x, s | x_{0}) = \frac{1}{\sqrt{s D}}\begin{cases}
   e^{-\sqrt{\frac{s}{D}}x_{0}}\hspace{1pt}\text{cosh}(\sqrt{\frac{s}{D}}x),  \hspace{12pt} x < x_{0},\\
  \text{cosh}(\sqrt{\frac{s}{D}}x_{0})e^{-\sqrt{\frac{s}{D}}x}, \hspace{10pt}  x > x_{0}.\\
 \end{cases}    
\end{equation}
Inverse Laplace transformation then gives
\begin{equation} \label{A:6}
C(x, t | x_{0}) = \frac{1}{\sqrt{4 \pi D t}} \Big( e^{- \frac{(x-x_{0})^2}{4 D t}}  + e^{- \frac{(x+x_{0})^2}{4 D t}} \Big).
\end{equation}
One can also get Eq. (\ref{A:6}) by using the method of images.

\section{SSLW approach applied to CTRW on networks} \label{Appendix:B} \setcounter{equation}{0}

Consider a gated process with internal dynamics that are described by Eq. (\ref{internal_propagator}) and a spatial process that is a CTRW on a general network. Let us denote explicitly the waiting time on the target site by $W_{0}$, and for simplicity assume that it is exponentially distributed with rate $\gamma$ (it can be easily generalized). Let the random
variable $\Vec{X}_1$ stand for the location of the particle following
the jump made after the waiting time $W_{0}$, namely, from the target to a different site (based on the connectivity of the network).

Equation 7 of Ref. \cite{scher2020unifying} gives the Laplace transform of the detection time, given that the walk started on the target in the non-reactive state $\mathbf{x}_{0} = (\mathbf{0},\text{NR}) := \mathbf{0}_{\text{NR}}$ (we set the origin on the target, $\mathbf{a} = \mathbf{0}$). Reproducing it here with the required slight change in notation:
\begin{align} \label{B:1}
&\tilde{T}_d\left(\mathbf{0}_{\textrm{NR}},s\right) = \\
&\frac{\pi_{\mathrm{R}}^{-1} K_{D}+\tilde{T}_{f}\left(\mathbf{X}_{1}, s\right)-\tilde{T}_{f}\left(\mathbf{X}_{1}, s+\lambda\right)}{\pi_{\mathrm{R}}^{-1}\left[(s / \gamma)+K_{D}+1\right]-K_{e q} \tilde{T}_{f}\left(\mathbf{X}_{\mathbf{1}}, s\right)-\tilde{T}_{f}\left(\mathbf{X}_{\mathbf{1}}, s+\lambda\right)}, \nonumber
\end{align}
where $ K_{D}= \frac{\beta}{\gamma}$ and $ K_{eq}= \frac{\alpha}{\beta}$ and $\tilde{T}_{f}\left(\mathbf{X}_{\mathbf{1}}, s\right)$ is the Laplace transform of the first-passage time to the origin starting from $\mathbf{X}_{\mathbf{1}}$.

We will now show how Eq. (\ref{B:1}) can be derived using the SSLW formalism (albeit by narrowing the range of validity to Markovian processes). Equation (\ref{SSLW_used}) for this specific initial condition reads
\begin{equation} \label{B:2}
 \tilde{T}_d(\mathbf{0}_{\textrm{NR}}, s)= \frac{\pi_{\textrm{R}}\tilde{C}\left(\mathbf{0}, s \mid \mathbf{0}\right) - \pi_{\textrm{R}} \tilde{C}\left(\mathbf{0}, s+\lambda \mid \mathbf{0}\right)  }{\pi_{\textrm{R}}\tilde{C}\left(\mathbf{0}, s \mid \mathbf{0}\right) + \pi_{\textrm{NR}} \tilde{C}\left(\mathbf{0}, s+\lambda \mid \mathbf{0}\right)}. 
\end{equation}

\noindent To proceed, we write the propagator $ C\left(\mathbf{0}, t \mid \mathbf{0}\right)$ explicitly in terms of the waiting time at the origin. The probability to not leave the origin at all during time $t$ is simply $e^{-\gamma t}$. Otherwise, the particle leaves at some point in the time interval $t' \in [0,t]$, with probability $\gamma e^{-\gamma t'}$. It leaves to $\Vec{X_1}$, the random location of the particle following this jump. Then we are looking for the remainder of the trajectory, from $\Vec{X_1}$ back to the target at the origin, in the remaining time $t-t'$. Concluding, we can re-write $C\left(\mathbf{0}, t \mid \mathbf{0}\right)$ as:
\begin{equation} \label{B:3}
    C\left(\mathbf{0}, t \mid \mathbf{0}\right) = e^{-\gamma t} + \int_{0}^{t} \gamma  e^{-\gamma t'}  C\left(\mathbf{0}, t-t' \mid \mathbf{X}_{1}\right) dt'.
\end{equation}
Laplace transforming we obtain
\begin{equation} \label{B:4}
    \tilde{C}\left(\mathbf{0}, s \mid \mathbf{0}\right) = \frac{1}{s+\gamma} + \frac{\gamma}{s+\gamma} \tilde{C}\left(\mathbf{0}, s \mid \mathbf{X}_{1}\right).
\end{equation}
Let $C \left(\mathbf{r}, t \mid \mathbf{r}_{0}\right)$ be the spatial propagator of a CTRW on the considered network. For such a walk, the following relation holds (\cite{klafter2011first}):
\begin{equation} \label{B:5}
   \tilde{C}(\mathbf{a},s \mid \mathbf{r}_{0})=\delta_{\mathbf{a}, \mathbf{r}_{0}}+\tilde{T}_{f}(s \mid \mathbf{r}_{0}) \tilde{C}(\mathbf{a},s \mid \mathbf{a})
\end{equation}
For the specific case above we have
\begin{equation} \label{B:6}
   \tilde{C}(\mathbf{0},s \mid \mathbf{X}_{1})=\tilde{T}_{f}(s \mid \mathbf{X}_{1}) \tilde{C}(\mathbf{0},s \mid \mathbf{0}).
\end{equation}

By plugging Eq. (\ref{B:6}) into Eq. (\ref{B:4}) we get:
\begin{equation} \label{B:7}
    \tilde{C}\left(\mathbf{0}, s \mid \mathbf{0}\right) = \frac{1}{s+\gamma} + \frac{\gamma}{s+\gamma} \tilde{T}_{f}(s \mid \mathbf{X}_{1}) \tilde{C}(\mathbf{0},s \mid \mathbf{0}),
\end{equation}
which can be re-written as
\begin{equation} \label{B:8}
    \tilde{C}\left(\mathbf{0}, s \mid \mathbf{0}\right) = \frac{1}{s+\gamma-\gamma \tilde{T}_{f}(s \mid \mathbf{X}_{1})}. 
\end{equation}
Plugging Eq. (\ref{B:8}) into Eq. (\ref{B:2}), after some algebra, we get Eq. (\ref{B:1}). A comprehensive analysis of this result can be found in Ref. \cite{scher2020unifying}.


\section{Laplace transform of the detection time distribution} \label{Appendix:C} \setcounter{equation}{0}

We start from Eq. (\ref{renewal_interval1}). Laplace transforming it we get
\begin{align} \label{C:1}
\tilde{T}_d(x_0,\sigma_0,s)&=\Braket{e^{-sT_d(x_0,\sigma_0) }} \\ &=\Braket{e^{-s\Big[ T_f(\rho,x_0) + I_f T_d(\rho, \text{NR})\Big]}}, \nonumber
\end{align}
which in turn gives
\begin{align} \label{C:2}
&\tilde{T}_d(x_0,\sigma_0,s) = \Braket{Q(\text{R},T_f(\rho, x_0 ) \mid \sigma_{0})e^{-sT_f(\rho, x_0 )}}
\\
 &+\Braket{Q(\text{NR},T_f(\rho, x_0 ) \mid \sigma_{0}) e^{-s\Big[T_f(\rho, x_0 ) + T_d(\rho, \text{NR})\Big]} }  \nonumber   \\
 &=\Braket{Q(\text{R},T_f(\rho, x_0 ) \mid \sigma_{0})e^{-sT_f(\rho, x_0 )}} \nonumber
 \\
 &+\Braket{Q(\text{NR},T_f(\rho, x_0 ) \mid \sigma_{0}) e^{-sT_f(\rho, x_0 )} }  \Braket{e^{-s T_d(\rho, \text{NR})}}. \nonumber
\end{align}
Setting $\sigma_{0}=\textrm{R}$ and using Eq. (\ref{internal_propagator}), we obtain
\begin{align} \label{C:3}
 &\tilde{T}_d(x_0,\textrm{R},s) = \Braket{\pi_{\text{R}}e^{-sT_f(\rho\mid x_0 )}  
+\pi_{\text{NR}}e^{-(s+\lambda)T_f(\rho\mid x_0 )}}
\\
&+\Braket{\pi_{\text{NR}}e^{-sT_f(\rho\mid x_0 ) }-\pi_{\text{NR}} e^{-(s+\lambda)T_f(\rho\mid x_0 )}}\Braket{e^{-s T_d(\rho, \text{NR})}}        \nonumber
 \\ \nonumber
 \\
 &=\pi_{\text{R}}\tilde{T}_f(\rho,s\mid x_0)+ \pi_{\text{NR}}\tilde{T}_f(\rho,s\mid x_0)\tilde{T}_d(\rho,\text{NR},s)  \nonumber
\\
&+ \pi_{\text{NR}}\tilde{T}_f(\rho,s+\lambda \mid x_0 ) - \pi_{\text{NR}}\tilde{T}_d(\rho,\text{NR},s)  \tilde{T}_f(\rho,s+\lambda \mid x_0) \nonumber
\\ \nonumber
\\
&=\pi_{\text{R}}\tilde{T}_f(\rho,s\mid x_0)+  \pi_{\text{NR}}\tilde{T}_f(\rho,s+\lambda \mid x_0) \nonumber
\\
&+ \Big[ \pi_{\text{NR}}\tilde{T}_f(\rho,s \mid x_0)- \pi_{\text{NR}} \tilde{T}_f(\rho,s+\lambda \mid x_0 )  \Big] \tilde{T}_d(\rho,\text{NR},s) . \nonumber
\end{align}
After repeating the same calculation for $\sigma_0=\textrm{NR}$ and $\sigma_0=\textrm{eq}$, we summarize the results in Eq. (\ref{distribution_1}).

Similarly, Laplace transforming Eq. (\ref{renewal_interval2})
\begin{align}  \label{C:4}
     &\Tilde{T}_d(\rho, \text{NR},s)=\int^{b}_{a} \Braket{ C(y,W_{\beta}\mid \rho) e^{- s W_{\beta}}  } dy \\
     & + \int^{\infty}_{-\infty} \Braket{ C(y,W_{\beta}\mid \rho) e^{- s W_{\beta}}  } \Braket{ e^{- s  T_d(y,\text{R})}  }\Big[\Theta_{-}(y)+\Theta_{+}(y)\Big]dy \nonumber
     \\ \nonumber
     \\
     &=\int^{b}_{a} \beta\tilde{C}(y,s+\beta\mid \rho)  dy \nonumber
     \\
     & + 
     \int^{\infty}_{-\infty} \beta\tilde{C}(y,s+\beta\mid \rho) \tilde{T}_d(y,\text{R},s) \Big[\Theta_{-}(y)+\Theta_{+}(y)\Big]dy \nonumber.
\end{align}
Let us denote the first integral term as
\begin{equation} \label{C:5}
 \tilde{\phi}_{\rho}(s) \equiv \beta\int^{b}_{a}\tilde{C}(y,s+\beta\mid \rho) dy \equiv  \int^{b}_{a} \tilde{\Phi}_{\rho}(s+\beta) dy. 
\end{equation}
Consider the second term in Eq. (\ref{C:4}). By plugging $\tilde{T}_d(y,\text{R},s)$ according to Eq. (\ref{C:3}) we obtain
\begin{align} \label{C:6}
\int^{\infty}_{-\infty} \tilde{\Phi}_{\rho}(s+\beta) &\tilde{T}_d(y,\text{R},s) \Big[\Theta_{-}(y)+\Theta_{+}(y)\Big]dy  
\\
&=  \tilde{\chi}_\rho(s) + \tilde{\psi}^{-}_{\rho} \tilde{T}_d(a,\text{NR},s)   + \tilde{\psi}^{+}_{\rho} \tilde{T}_d(b,\text{NR},s), \nonumber
\end{align}
where
\begin{align} \label{C:7}
&\tilde{\chi}_\rho(s) \equiv \int^{\infty}_{-\infty} \tilde{\Phi}_{\rho}(s+\beta) \times \\ &\Big[\pi_{\text{R}}\tilde{T}_f(\iota_{\pm},s \mid y) + \pi_{\text{NR}}\tilde{T}_f(\iota_{\pm},s+\lambda \mid y) \Big] 
 \Big[\Theta_{-}(y)+\Theta_{+}(y)\Big]dy ,  \nonumber
\end{align}
and
\begin{align} \label{C:8}
\tilde{\psi}^{\pm}_{\rho}(s)  \equiv &  \int^{\infty}_{-\infty} \tilde{\Phi}_{\rho}(s+\beta) \times
\\
& \pi_{\text{NR}} \Big[ \tilde{T}_f(\iota_\pm,s \mid y) - \tilde{T}_f(\iota_\pm,s+\lambda \mid y) \Big]  \Theta_{\pm}(y)dy,   \nonumber
\end{align}
where $\iota_- = a$ and $\iota_+ = b$. Overall we have
\begin{align} \label{C:9}
     \Tilde{T}_d(\rho, \text{NR},s) &= \tilde{\Phi}_{\rho}(s+\beta) + \tilde{\chi}_{\rho}(s) \\
     &+ \tilde{\psi}^{-}_{\rho} \tilde{T}_d(a,\text{NR},s)   + \tilde{\psi}^{+}_{\rho} \tilde{T}_d(b,\text{NR},s) . \nonumber
\end{align}
Note that Eq. (\ref{C:9}) represents a linear system of two equations with the two unknowns $\Tilde{T}_d(a, \text{NR},s)$ and $ \Tilde{T}_d(b, \text{NR},s)$. Solving we get:
\begin{equation} \label{C:10}
 \Tilde{T}_d(a,\text{NR},s)    = 
\frac{(\tilde{\phi}_a + \tilde{\chi}_a)(1- \tilde{\psi}^{+}_{b}) +  \tilde{\psi}^{+}_{a} (\tilde{\phi}_b + \tilde{\chi}_b)}{1 - \tilde{\psi}^{-}_{a} -\tilde{\psi}^{+}_{b} + \tilde{\psi}^{-}_{a}\tilde{\psi}^{+}_{b}  - \tilde{\psi}^{-}_{b}\tilde{\psi}^{+}_{a} } , 
\end{equation}
and
\begin{equation} \label{C:11}
 \Tilde{T}_d(b,\text{NR},s)    = 
\frac{ (\tilde{\phi}_b + \tilde{\chi}_b)(1- \tilde{\psi}^{-}_{a})  + \tilde{\psi}^{-}_{b}(\tilde{\phi}_a + \tilde{\chi}_a)  }{1 - \tilde{\psi}^{-}_{a} -\tilde{\psi}^{+}_{b} + \tilde{\psi}^{-}_{a}\tilde{\psi}^{+}_{b}  - \tilde{\psi}^{-}_{b}\tilde{\psi}^{+}_{a} } . 
\end{equation}
For the symmetric case in which the spatial dynamics and the boundary conditions to the left and the right of the target center are the same (e.g., diffusion on the infinite line) Eqs. (\ref{C:10}) and (\ref{C:11}) are equal and simplify considerably ($\phi_a = \phi_b := \phi$, $\chi_a = \chi_b := \chi$ and $\psi^{\pm}_{a} = \psi^{\pm}_{b}:=\psi^{\pm}$)
\begin{equation} \label{C:12}
 \Tilde{T}_d(a, \text{NR},s) = \Tilde{T}_d(b, \text{NR},s)   = \frac{\tilde{\phi} + \tilde{\chi}}{1 - \tilde{\psi}^{-} -\tilde{\psi}^{+} }.
\end{equation}

\section{Long time asymptotics -- Inheritance of power-laws} \label{Appendix:D} 
\setcounter{equation}{0}

Assume the symmetric case in which the spatial dynamics and the boundary conditions to the left and the right of the target center are the same. Further assume that the underlying ungated process has an asymptotic power-law behavior of the form of {$\tilde{T}_f(\rho,s \mid y) \simeq1- (\tau_f s)^{\theta}$ (for $s\ll1$), where $0<\theta<1$ and $\tau_f>0$. Note that $\tau_f$ is a function of $\abs{y-\rho}$. Taking the limit $s \to 0$ and plugging this form into Eq. (\ref{distribution_4}) we obtain 
\begin{equation} \label{D:1}
 \Tilde{T}_d(\rho, \text{NR}, s) \simeq
\frac{ K_{eq} A - B s^{\theta} }{ K_{eq} A' + K_{eq} B s^{\theta} },  
\end{equation}
where $K_{e q}=\alpha / \beta$ and 
\begin{align} \label{D:2}
  & A = \int^{a}_{-\infty} \tilde{\Phi}_{a}(\beta) \tilde{T}_f(a,\lambda \mid y) dy
   +\int^{\infty}_{b} \tilde{\Phi}_{a}(\beta)  \tilde{T}_f(b,\lambda \mid y) dy + \nonumber
  \\
  &  \frac{\pi^{-1}_{\textrm{R}}\int^{b}_{a} \tilde{\Phi}_{a}(\beta)dy+ 
     \int^{a}_{-\infty} \tilde{\Phi}_{a}(\beta)  dy +\int^{\infty}_{b} \tilde{\Phi}_{a}(\beta) dy }{K_{eq}} , 
\end{align}

\begin{equation}  \label{D:3}
\begin{array}{ll}
  B =   \int^{a}_{-\infty} \tilde{\Phi}_{a}(\beta) \tau_f^{\theta}(y) dy +\int^{\infty}_{b} \tilde{\Phi}_{a}(\beta) \tau^{\theta}_f(y) dy
\end{array}
\end{equation}
and
\begin{align} \label{D:4}
  &A' = \pi^{-1}_{\text{NR}}
      -\int^{a}_{-\infty} \tilde{\Phi}_{a}(\beta)dy   - \int^{\infty}_{b} \tilde{\Phi}_{a}(\beta) dy 
     \\
  &+ \int^{a}_{-\infty} \tilde{\Phi}_{a}(\beta) \tilde{T}_f(a,\lambda\mid y) dy
+\int^{\infty}_{b} \tilde{\Phi}_{a}(\beta) \tilde{T}_f(b,\lambda \mid y) dy \nonumber
  .
\end{align}
The numerator of the second row in Eq. (\ref{D:2}) can be rewritten as
\begin{align} \label{D:5}
& \pi^{-1}_{\textrm{R}}\int^{b}_{a} \tilde{\Phi}_{a}(\beta)dy+  
     \int^{a}_{-\infty} \tilde{\Phi}_{a}(\beta)  dy +\int^{\infty}_{b} \tilde{\Phi}_{a}(\beta) dy   \nonumber
 \\
& =  \int^{\infty}_{-\infty} \tilde{\Phi}_{a}(\beta) dy + ( \pi^{-1}_{\textrm{R}}-1) \int^{b}_{a} \tilde{\Phi}_{a}(\beta) dy                         \nonumber
 \\
& = \int^{\infty}_{-\infty} \tilde{\Phi}_{a}(\beta) dy + K_{eq} \int^{b}_{a} \tilde{\Phi}_{a}(\beta)dy  
 \\
&= 1 + K_{eq} \int^{b}_{a} \tilde{\Phi}_{a}(\beta)dy, \nonumber
\end{align}
where in the last transition we recall that $\tilde{\Phi}_{a}(\beta)=\beta \tilde{C}(y,\beta\mid a)$ and note that $\int^{\infty}_{-\infty} \tilde{C}(y,\beta\mid a) dy = \int^{\infty}_{-\infty} \int^{\infty}_{0} {C}(y,t \mid a) e^{-\beta t} dt dy  = \int^{\infty}_{0} \int^{\infty}_{-\infty}  {C}(y,t \mid a) e^{-\beta t} dy dt  = \int^{\infty}_{0} e^{-\beta t} dy dt = \beta^{-1}$. Thus, $A$ in Eq. (\ref{D:2}) can be rewritten as
\begin{align}  \label{D:6}
  &A = K^{-1}_{eq} +  \int^{b}_{a} \tilde{\Phi}_{a}(\beta)dy  +
  \\
  &\int^{a}_{-\infty} \tilde{\Phi}_{a}(\beta) \tilde{T}_f(a,\lambda\mid y) dy +\int^{\infty}_{b} \tilde{\Phi}_{a}(\beta)  \tilde{T}_f(b,\lambda\mid y) dy  . \nonumber
\end{align}
By noting that $\pi^{-1}_{\text{NR}} - 1 = K^{-1}_{eq}$ it is easy to see that
\begin{equation} \label{D:7}
  A = A'.
\end{equation}
By algebraic manipulations, and given Eq. (\ref{D:7}), Eq. (\ref{D:1}) can be rewritten as 
\begin{equation} \label{D:8}
 \Tilde{T}_d(\rho, \text{NR}, s) \simeq
\frac{K_{eq} A \Big(1-\frac{B s^{\theta}}{K_{eq} A}\Big)  }{K_{eq} A' \Big(1+\frac{ B s^{\theta}}{A'}\Big)  } = \frac{1-\frac{B s^{\theta}}{K_{eq} A}  }{1+\frac{B s^{\theta}}{A}  },
\end{equation}
which is equivalent to
\begin{equation} \label{D:9}
 \Tilde{T}_d(\rho, \text{NR}, s) \simeq
 \Big(1-\frac{B s^{\theta}}{K_{eq} A}\Big) \frac{1}{1+\frac{B s^{\theta}}{A} }.    
\end{equation}
Expanding the fraction on the right, we obtain
\begin{equation} \label{D:10}
 \Tilde{T}_d(\rho, \text{NR}, s) \simeq
\Big(1-\frac{B s^{\theta}}{K_{eq}A}\Big) \Big(1-\frac{B s^{\theta}}{A}\Big) .    
\end{equation}
Now by carefully multiplying these terms and neglecting higher order products we get 
\begin{equation} \label{D:11}
 \Tilde{T}_d(\rho, \text{NR}, s) \simeq
 1-\frac{B(1 + K_{eq})}{K_{eq}A} s^{\theta} =  1-\pi^{-1}_{\text{NR}}\frac{B}{A} s^{\theta}.
\end{equation}
Lastly, by an indirect application of the Tauberian theorem (see pp. 43-45 in \cite{klafter2011first}) we obtain
\begin{equation} \label{D:12}
 f_d(t,\rho, \text{NR}) \simeq
  \frac{\theta }{\Gamma(1-\theta)} \frac{\left(\pi^{-1}_{\text{NR}}\frac{B}{A}\right)}{t^{1+\theta}}.
\end{equation}
 
\section{Transient behaviour under high crypticity} \label{Appendix:E} 
\setcounter{equation}{0}

\subsection{Ungated first-passage distributions with power-law tails}

Recall Eq. (\ref{D:1}) (we keep working under the same assumptions of Appendix \ref{Appendix:D})
\begin{equation} \label{E:1}
  \Tilde{T}_d(\rho, \text{NR}, s) \simeq
\frac{K_{eq} A - B s^{\theta} }{K_{eq} A + K_{eq} B s^{\theta} },  
\end{equation}
where A and B are defined as in the previous section and we recall that $A=A'$ (Eq. (\ref{D:7})). By taking the limit $K_{eq} \gg 1$, we obtain
\begin{equation} \label{E:2}
   \Tilde{T}_d(\rho, \text{NR}, s) \simeq \frac{K_{eq} A }{K_{eq} A + K_{eq} B s^{\theta} } = \frac{1 }{1 + \frac{B}{A}  s^{\theta} } .  
\end{equation}
To obtain a transient pre-asymptotic behaviour, we require the existence of $s$ values such that $\beta > s \gg (\frac{A}{B})^{1/\theta}$. In this case, $\frac{B}{A}s^\theta \gg 1$}, and we have
\begin{equation} \label{E:3}
\Tilde{T}_d(\rho, \text{NR}, s) \simeq \frac{1}{\frac{B}{A}  s^{\theta} } .  
\end{equation}
Applying the Tauberian theorem gives
\begin{equation} \label{E:4}
 f_d(t,\rho, \text{NR}) \simeq \frac{1}{\Gamma(\theta)}\frac{A}{B}  t^{\theta-1}, 
\end{equation}
which is a transient regime with a different power law than the asymptotic power law. Furthermore, we can exactly determine this power, and the pre-exponential factor. Thus, to guarantee the existence of the pre-asymptotic behaviour in Eq. (\ref{E:4}), we require $\frac{B}{A}\beta^\theta \gg 1$. We note that if this requirement is not fulfilled, the transition from Eq. (\ref{E:2}) to Eq. (\ref{E:3}) is invalid. Expanding  Eq. (\ref{E:2}) (in the limit $s \to 0$) we then obtain 
\begin{equation} \label{E:5}
 \Tilde{T}_d(\rho, \text{NR}, s) \simeq
 1-\frac{B}{A} s^{\theta}, 
\end{equation}
which, as expected, is the $K_{eq} \gg 1$ limit of Eq. (\ref{D:11}). This means that there is no transient regime before the asymptotic regime kicks in.

To understand the meaning of the additional requirement $\frac{B}{A} \beta^{\theta} \gg 1$, we focus on propagators whose Laplace transform has a scaling form 
\begin{equation} \label{E:scaling}
\tilde{C}(y,s\mid a) \simeq s^{-\theta} H(\abs{y-a}s^\theta), 
\end{equation}
where $H$ is the scaling function. Note that this form generalizes the one displayed by one-dimensional free-diffusion, $ \tilde{C}(x, s | x_{0}) =\frac{1}{\sqrt{4Ds}} e^{-\sqrt{\frac{s}{D}} \abs{x-x_{0}}}$. It is then easy to show that under this assumption $B\beta^\theta \sim O(1)$, i.e., this product does not scale with $\beta$. Indeed, the assumed scaling sets a relation between length and time scales in our problem: length $\sim$ time$^\theta$. This means that
\begin{align}  \label{E:5a}
   & B =   \int^{a}_{-\infty} \tilde{\Phi}_{a}(\beta) \tau_f^{\theta}(y) dy +\int^{\infty}_{b} \tilde{\Phi}_{a}(\beta) \tau^{\theta}_f(y) dy 
  \\
  & \sim \int^{a}_{-\infty} \tilde{\Phi}_{a}(\beta) \abs{y-a} dy +\int^{\infty}_{b} \tilde{\Phi}_{a}(\beta) \abs{y-b} dy \sim \beta^{-\theta}, \nonumber
\end{align}
where we have noted that $\beta^{-1}$ sets the time scale, and hence the typical length scale that is set by $\tilde{\Phi}_{\rho}(\beta)=\beta \tilde{C}(y,\beta\mid \rho)$ (recall that this function is normalized to unity over space) scales like $\beta^{-\theta}$. Hence, $B\beta^\theta$ is order one, which means that to satisfy $\frac{B}{A} \beta^{\theta} \gg 1$, one only needs $A\ll1$.

Equation (\ref{D:6}) gives $A$ as a sum of four terms. The first term is small since we assumed $ K_{eq} \gg 1$. To make the second term small, we require $\int^{b}_{a}\tilde{\Phi}_{a}(\beta) = \beta\int^{b}_{a} \tilde{C}(y,\beta\mid a)dy \ll 1$, which can also be written as $\int^{b}_{a}\tilde{C}(y,\beta\mid a)dy \ll \beta^{-1}$. We anticipate that this condition can be fulfilled by taking $\beta$ to be small. In this limit, we can approximate
\begin{equation} \label{E:6}
 \beta\int^{b}_{a} \tilde{C}(y,\beta\mid a)dy \simeq (b-a)  \beta\tilde{C}(a,\beta\mid a), 
\end{equation}
since the probability of being at different points inside the interval is practically the same at long times, which corresponds here to the small $\beta$ limit of the Laplace transform $\tilde{C}(y,\beta\mid a)$. To find how the expression in Eq. (\ref{E:6}) scales with $\beta$, we utilize a well established relation between $\tilde{C}(y,s\mid a)$, $\tilde{C}(a,s\mid a)$ and $\tilde{T}_f(y,s\mid a)$:
\begin{align} \label{E:7}
 \tilde{T}_f(y,s\mid a) = & \frac{\tilde{C}(y,s\mid a)}{\tilde{C}(a,s\mid a)}
 \\
 &\approx \frac{\tilde{C}(a,s\mid a) + \abs{y-a} \frac{d\tilde{C}(y,s\mid a)}{dy} \Big|_{y=a}}{\tilde{C}(a,s\mid a)} . \nonumber
\end{align}
We thus see that for $\tilde{T}_f(\rho,s \mid y) \simeq 1- (\tau_f s)^{\theta}$
, where $0<\theta<1$ and $\tau_f>0$, we have
\begin{equation} \label{E:8}
 \tilde{C}(a,s\mid a) \simeq \left(\frac{\abs{y-a} \frac{d\tilde{C}(y,s\mid a)}{dy}}{\tau_f^{\theta}}\right)_{y=a} s^{-\theta} \equiv (\tau_r s)^{-\theta}.
\end{equation}

Recalling the scaling form in  Eq. (\ref{E:scaling}), we see that the derivative with respect to $y$, evaluated at $y=a$, is independent of $s$. Thus, $\tilde{C}(a,s\mid a)$ in Eq. (\ref{E:8}) scales like $s^{-\theta}$, with a prefactor $\tau_r \equiv \left(\abs{y-a} \frac{d\tilde{C}(y,s\mid a)}{dy} \tau_f^{-\theta}(\abs{y-a})\right)^{-1/\theta} \Big|_{y=a}$, where we have explicitly recalled that $\tau_f$ is a function of $\abs{y-a}$. Plugging Eq. (\ref{E:8}) back into Eq. (\ref{E:6}) we obtain
\begin{equation} \label{E:9}
 \beta\int^{b}_{a} \tilde{C}(y,\beta\mid a)dy \simeq \frac{b-a}{\tau_r^\theta} \beta^{1-\theta},
\end{equation}
which is indeed small for small enough $\beta$ since $0<\theta<1$. More precisely, the additional requirement $\beta\int^{b}_{a} \tilde{C}(y,\beta\mid a)dy \ll 1$ translates to 
\begin{equation} \label{E:10}
 \frac{b-a}{\tau_r^\theta} \ll \beta^{\theta-1} ,
\end{equation}

\noindent which asserts that the second term in Eq. (\ref{D:6}) is small. 

We are left with the last two terms in Eq. (\ref{D:6}). To analyze their scaling with $\beta$, we once again observe that the typical length scale that is set by $\tilde{\Phi}_{\rho}(\beta)=\beta \tilde{C}(y,\beta\mid \rho)$ scales like $\sim \beta^{-\theta}$. Hence, for the third term 
\begin{align}  \label{E11}
  &\int^{a}_{-\infty} \tilde{\Phi}_{a}(\beta) \tilde{T}_f(a,\lambda\mid y) dy \sim  \tilde{T}_f(a,\lambda\mid \bar{y}), 
\end{align}
with $\bar{y}$ set such that $|\bar{y}-a|\sim \beta^{-\theta}$. We now note that $\tilde{T}_f(a,\lambda\mid \bar{y})$ is a Laplace transform, which in the limit $\lambda \tau_f(|\bar{y}-a|) \ll 1$ can be approximated by $\tilde{T}_f(a,\lambda \mid \bar{y}) \simeq 1- (\tau_f \lambda)^{\theta}$. In the other limit, i.e., when  
\begin{align}  \label{E12}
  &\lambda \tau_f(|\bar{y}-a|) \gg 1,
\end{align}
$\tilde{T}_f(a,\lambda\mid \bar{y})$ rapidly tends to zero. To see that we are in the limit specified by Eq. (\ref{E12}), we simply observe that 
\begin{align}  \label{E13}
  \lambda \tau_f(|\bar{y}-a|) \sim \lambda |\bar{y}-a|^{1/\theta} \sim \lambda/\beta=K_{eq} +1\gg1, 
\end{align}
\noindent since we started by assuming $K_{eq} \gg 1$. As the analysis of the fourth term yields the same result, we conclude that the additional condition in Eq. (\ref{E:10}) is sufficient to guarantee the pre-asymptotic behaviour in Eq. (\ref{E:4}).



\subsection{Implications to gated CTRW
on networks}
We return to Eq. (\ref{B:1}) adapted from Ref. \cite{scher2020unifying}, where we recall that $\gamma$ is the rate of escape from the target site, $K_D = \beta/\gamma$.  Equation. (\ref{B:1}) can be rewritten as
\begin{align} \label{E:18}
    &\widetilde{T}_d(\mathbf{0}_{\textrm{NR}}, s) =
    \\ 
    &\frac{ K_D + \pi_{\text{R}}\big[\widetilde{T}_f(\Vec{X}_1,s)-\widetilde{T}_f(\Vec{X}_1,s+\lambda)\big]}{\frac{s}{\gamma} + K_D +1 -\pi_{\text{R}} K_{eq}\widetilde{T}_f(\Vec{X}_1,s)-\pi_{\text{R}}\widetilde{T}_f(\Vec{X}_1,s+\lambda)} \nonumber
\end{align}
Taking the limit $K_{eq} \gg 1$ (such that $\pi_{\text{R}} \ll 1$ and $K_{eq} \pi_{\text{R}} \approx 1$), we obtain
\begin{equation} \label{E:19}
    \widetilde{T}_d(\mathbf{0}_{\textrm{NR}}, s) \simeq \frac{ K_D }{\frac{s}{\gamma} + K_D +1 -\widetilde{T}_f(\Vec{X}_1,s)}
\end{equation}  

Next we take the limit $s \to 0$, while assuming that in this limit $\widetilde{T}_f(\Vec{X}_1,s) \simeq 1 - (\tau_f s)^\theta$. If $K_D \ll 1$, which is equivalent to $\gamma^{-1} \ll \beta^{-1}$, we obtain
\begin{equation} \label{E:20}
    \widetilde{T}(\mathbf{0}_{\textrm{NR}}, s) \simeq \frac{ K_D }{  (\tau_f s)^\theta
    },
\end{equation}
By using the Tauberian theorem the inverse Laplace transform gives
\begin{equation} \label{E:21}
 f_d(t,\rho, \text{NR}) \simeq \frac{1}{\Gamma(\theta)}\frac{K_D}{\tau_f^\theta}  t^{\theta-1},
\end{equation}
which is a transient regime with a different power law than the asymptotic power law. Furthermore, we can exactly determine this power, and the pre-exponential factor.

If $K_D \not \ll 1$, Eq. (\ref{E:20}) is not valid, and we instead have
\begin{equation} \label{E:22}
\widetilde{T}_d(\mathbf{0}_{\textrm{NR}}, s) \simeq \frac{ 1 }{1 + \frac{(\tau_f s)^\theta}{K_D}},
\end{equation}
which, in the limit $s \to 0$, can be expanded as
 \begin{equation} \label{E:23}
\widetilde{T}_d(\mathbf{0}_{\textrm{NR}}, s) \simeq 1 - \frac{(\tau_f s)^\theta}{K_D},
\end{equation}
which is exactly the asymptotic behaviour reported in Eq. 8 of Ref. \cite{scher2020unifying} under the condition $K_{eq} \gg 1$. Thus, when $K_D \not \ll 1$, we do not have a transient regime before the asymptotic behaviour kicks in.\\

\section{Diffusion in an interval with two reflecting boundaries} \label{Appendix:F} \setcounter{equation}{0}

\subsection{The conserved propagator}

The propagator obeys the diffusion equation. The initial condition is $C(x, t=0 | x_{0})=\delta\left(x-x_{0}\right)$, and we require two Neumann boundary conditions $\frac{\partial C(x, t | x_{0})}{\partial x} \big|_{x=0} = 0$ and $\frac{\partial C(x, t | x_{0})}{\partial x} \big|_{x=L} = 0$. 

Laplace transforming the diffusion equation we get
\begin{equation} \label{F:1}
   s \tilde{C}(x, s | x_{0}) = D \frac{d^2 \tilde{C}(x, s | x_{0})}{dx^2}, 
\end{equation}
which is a second-order, linear, homogeneous differential equation.  It has a general solution
\begin{equation} \label{F:2} 
 \tilde{C}(x, s | x_{0}) =
 \begin{cases}
 \tilde{C}_{<} = A_{1}(s) \mathrm{e}^{\alpha_{+} x}+B_{1}(s) \mathrm{e}^{\alpha_{-} x},  \hspace{3pt} x < x_{0},\\
 \tilde{C}_{>} =A_{2}(s) \mathrm{e}^{\alpha_{+} x}+B_{2}(s) \mathrm{e}^{\alpha_{-} x}, \hspace{3pt}  x > x_{0},\\
 \end{cases}    
\end{equation}

\noindent where $\alpha_{\pm}=\pm\sqrt{\frac{s}{D}}$.

Similarly, Laplace transforming the boundary conditions we obtain 
\begin{equation} \label{F:3}
   \frac{d \tilde{C}(x, s | x_{0})}{d x} \big|_{x=0} = 0,
\end{equation}
and
\begin{equation}  \label{F:4}
    \frac{d \tilde{C}(x, s | x_{0})}{d x} \big|_{x=L} = 0.
\end{equation}

Finally, the initial condition is translated to two matching conditions at the initial position of the particle, one for the continuity of the Laplace transform of the probability density
\begin{equation}  \label{F:5}
     \tilde{C}_{+}(x, s | x_{0}) = \tilde{C}_{-}(x, s | x_{0}),  
\end{equation}
and one for the Laplace transform of the fluxes 
\begin{equation}  \label{F:6}
     -1=D\Big[\frac{d \tilde{C}_{+}(x, s | x_{0})}{d x}\big|_{x=0}  - \frac{d \tilde{C}_{-}(x, s | x_{0})}{d x}\big|_{x=0}\Big], 
\end{equation}
which is obtained by integrating both sides of the Laplace transformed diffusion equation (Eq.~(\ref{F:1})) over an infinitesimally small interval surrounding the initial position. Note that the $1$ on the left-hand side comes from the Laplace transform of the delta function initial condition. 

Imposing these conditions produces a system of four equations with four unknowns, from which we obtain
\begin{equation} \label{F:7}
\begin{cases}
  A_1(s) = \frac{\text{csch}\left(L \sqrt{\frac{s}{D}}\right) \cosh \left(\sqrt{\frac{s}{D}}
   \left(L-x_0\right)\right)}{2 \sqrt{D s}},
  \\
  B_1(s) = A_1(s),
  \\
  A_2(s) = \frac{\left[\coth \left(L \sqrt{\frac{s}{D}}\right)-1\right] \cosh \left(x_0 \sqrt{\frac{s}{D}}\right)}{2
   \sqrt{D s}},
  \\
  B_2(s) = \frac{\left[\coth \left(L \sqrt{\frac{s}{D}}\right)+1\right] \cosh \left(x_0 \sqrt{\frac{s}{D}}\right)}{2
   \sqrt{D s}}.
\end{cases}
\end{equation}

\subsection{First passage time to an absorbing boundary at $a$}

We now assume that $x_0 \in [0,a]$, where $0<a \leq b<L$, and repeat the exact same procedure but with different boundary conditions, $\frac{\partial C(x, t | x_{0})}{\partial x} \big|_{x=0} = 0$ and $C(a | x_{0} ) = 0$. The Laplace transformed boundary conditions are
\begin{equation}  \label{F:8}
     \tilde{C}_{>}(a,s) = 0,
\end{equation}
and
\begin{equation}  \label{F:9}
      \frac{d \tilde{C}_{<}(x,s)}{d x}\big|_{x=0}  = 0.
\end{equation}

\noindent Solving, we get
\begin{equation}  \label{F:10}
\begin{cases}
  A_1(s) = \frac{\text{sech}\left(a \sqrt{\frac{s}{D}}\right) \sinh \left(\left(a-x_0\right)
   \sqrt{\frac{s}{D}}\right)}{2 \sqrt{D s}},  \\
  
  B_1(s) =  A_1(s),
  \\
  A_2(s) = \frac{\left[\tanh \left(a \sqrt{\frac{s}{D}}\right)-1\right] \cosh \left(x_0
   \sqrt{\frac{s}{D}}\right)}{2 \sqrt{D s}},
  \\
  B_2(s) = \frac{\left[\tanh \left(a \sqrt{\frac{s}{D}}\right)+1\right] \cosh \left(x_0
   \sqrt{\frac{s}{D}}\right)}{2 \sqrt{D s}}.
  
\end{cases}
\end{equation}

The Laplace transform first-passage probability is given by
\begin{align}  \label{F:11}
   \tilde{T}_f(x=a, s| x_{0}) =& -\left.D \frac{d \tilde{C}_{>}(x, s)}{d x}\right|_{x=a} 
   \\
   &=\text{sech}\left(a \sqrt{\frac{s}{D}}\right) \cosh \left(x_0 \sqrt{\frac{s}{D}}\right) \nonumber
\end{align}

\noindent The Laplace transform is a moment generating function, the mean first-passage time is given by
\begin{align}   \label{F:12}
   \braket{T_f (a| x_{0})}= & -\frac{d \tilde{T}_f(a, s)}{d s}|_{s=0} 
   \\
   &=\frac{a^2-x_0^2}{2 D}. \nonumber
\end{align}


\section{Diffusion with drift on the positive half line with a reflecting boundary condition at the origin} \label{Appendix:H}
\setcounter{equation}{0}

\subsection{The conserved propagator}

We compute the propagator for the diffusion equation with drift, Eq. (\ref{drift}), with the initial condition $C(x, t=0 | x_{0})=\delta\left(x-x_{0}\right)$ and boundary conditions $\frac{\partial C(x, t | x_{0})}{\partial x} \big|_{x=0} = 0$ and $C(x \to \infty, t  | x_{0} ) = 0$. The Drift velocity is $v>0$ and its direction is towards the reflecting boundary at zero.

Laplace transforming the diffusion equation we have
\begin{equation} \label{H:1}
    s \tilde{C}(x, s  | x_{0} )-\delta\left(x-x_{0}\right)=D \frac{d^{2}\tilde{C}(x, s | x_{0})}{dx^2} + v \frac{d\tilde{C}(x, s | x_{0})}{dx}.
\end{equation}
This is a second-order, linear, non-homogeneous differential equation.
It has general spatial coordinate-dependent solution
\begin{equation} \label{H:2} 
 \tilde{C}(x, s | x_{0}) =
 \begin{cases}
 \tilde{C}_{<} = A_{1}(s) \mathrm{e}^{\alpha_{+} x}+B_{1}(s) \mathrm{e}^{\alpha_{-} x},  \hspace{1pt} x < x_{0},\\
 \tilde{C}_{>} =A_{2}(s) \mathrm{e}^{\alpha_{+} x}+B_{2}(s) \mathrm{e}^{\alpha_{-} x}, \hspace{1pt}  x > x_{0},\\
 \end{cases}    
\end{equation}
where $\alpha_{\pm}=\frac{1}{2 D}\left[-v \pm \sqrt{v^{2}+4 D s}\right]$.

At $x_{0}$ we impose two matching conditions, one for the propagator
\begin{equation} \label{H:3}  
     \tilde{C}_{>}(x_{0},s) =  \tilde{C}_{<}(x_{0},s),  
\end{equation}
and one for the fluxes (by integrating both sides of the transformed diffusion equation around an infinitesimally small interval)
\begin{equation} \label{H:4}
-1=D\Big[\frac{d \tilde{C}_{>}}{d x}\big|_{x=x_{0}}  - \frac{d \tilde{C}_{<}}{d x}\big|_{x=x_{0}}\Big]
\end{equation}
where the $1$ in the left-hand side comes from the Laplace transform of the delta function initial condition.

Finally, the transformed boundary conditions are
\begin{equation} \label{H:5}
     \tilde{C}_{>}(x \to \infty,s) = 0,
\end{equation}
and
\begin{equation} \label{H:6}
      v\tilde{C}_{<}(0,s)+D\frac{d \tilde{C}_{<}(x,s)}{d x}\big|_{x=0}  = 0.
\end{equation}
Solving, we get
\begin{equation} \label{H:7}
\begin{cases}
  A_1(s) = \frac{e^{-x_0 \alpha_{+}}}{D\left(\alpha_{+}-\alpha_{-}\right)},
  \\
  B_1(s) = \frac{(v+D\alpha_{+})e^{-x_0 \alpha_{+}}}{ D(v+D\alpha_{-})(\alpha_{-}-\alpha_{+})},
  \\
  A_2(s) = 0,
  \\
  B_2(s) = B_1(s)-\frac{e^{-x_0 \alpha_{-}}}{D\left(\alpha_{+}-\alpha_{-}\right)}.
\end{cases}
\end{equation}

\subsection{First-Passage Time to an absorbing boundary at $a$}

We now repeat the exact same process but with different boundary conditions, $\frac{\partial C(x, t | x_{0})}{\partial x} \big|_{x=0} = 0$ and $C(a | x_{0} ) = 0$. The Laplace transformed boundary conditions are
\begin{equation}  \label{H:8}
     \tilde{C}_{>}(a,s) = 0,
\end{equation}
and
\begin{equation}  \label{H:9}
      \frac{d \tilde{C}_{<}(x,s)}{d x}\big|_{x=0}  = 0.
\end{equation}

Solving, we get
\begin{equation}  \label{H:10}
\begin{cases}
  A_1(s) = \frac{\alpha_{-} e^{-x_{0}\left(\alpha_{-}+\alpha_{+}\right)}\left(e^{x_{0} \alpha_{-}+L \alpha_{+}}-e^{L \alpha_{-}+x_{0} \alpha_{+}}\right)}{D\left(\alpha_{-}-\alpha_{+}\right)\left(\alpha_{+} e^{L \alpha_{-}}-\alpha_{-} e^{L \alpha_{+}}\right)},  \\
  
  B_1(s) = \frac{e^{L \alpha_{-}}\left(\alpha_{+} e^{-x_{0} \alpha_{+}}-\alpha_{-} e^{-x_{0} \alpha_{-}}\right)}{D\left(\alpha_{-}-\alpha_{+}\right)\left(\alpha_{+} e^{L \alpha_{-}}-\alpha_{-} e^{L \alpha_{+}}\right)},
  \\
  A_2(s) = \frac{\alpha_{+} e^{-x_{0}\left(\alpha_{-}+\alpha_{+}\right)}\left(e^{L \alpha_{-}+x_{0} \alpha_{+}}-e^{x_{0} \alpha_{-}+L \alpha_{+}}\right)}{D\left(\alpha_{-}-\alpha_{+}\right)\left(\alpha_{+} e^{L \alpha_{-}}-\alpha_{-} e^{L \alpha_{+}}\right)},
  \\
  B_2(s) = \frac{e^{L \alpha_{+}}\left(\alpha_{+} e^{-x_{0} \alpha_{+}}-\alpha_{-} e^{-x_{0} \alpha_{-}}\right)}{D\left(\alpha_{-}-\alpha_{+}\right)\left(\alpha_{+} e^{L \alpha_{-}}-\alpha_{-} e^{L \alpha_{+}}\right)}.
\end{cases}
\end{equation}

The Laplace transform first-passage probability is given by
\begin{align}  \label{H:11}
   \tilde{T}_f(x=a, s) =& -\left.D \frac{d \tilde{C}_{>}(x, s)}{d x}\right|_{x=a} 
   \\
   &=\frac{e^{\left(a-x_{0}\right)\left(\alpha_{-}+\alpha_{+}\right)}\left(\alpha_{-} e^{x_{0} \alpha_{+}}-\alpha_{+} e^{x_{0} \alpha_{-}}\right)}{\alpha_{-} e^{a \alpha_{+}}-\alpha_{+} e^{a \alpha_{-}}}. \nonumber
\end{align}
The Laplace transform is a moment generating function, the mean first-passage time is given by
\begin{align}   \label{H:12}
   \braket{T_f (a| x_{0})}= & -\frac{d \tilde{T}_f(a, s)}{d s}|_{s=0} 
   \\
   &=\frac{D \left(-e^{-\frac{a v}{D}}+e^{\frac{(x_{0}-2 a) v}{D}}\right)+e^{\frac{(x_{0}-a) v}{D}}(a-x_{0}) v}{v^{2}}. \nonumber
\end{align}

\section{Diffusion in a logarithmic potential on the positive half line  with a reflecting boundary condition at the origin} \label{Appendix:G} \setcounter{equation}{0}

\subsection{The conserved propagator}

We compute the propagator for the Smoluchowski equation with a logarithmic potential (Eq. (\ref{log1})), the initial condition $C(x, t=0 | x_{0})=\delta\left(x-x_{0}\right)$ and the boundary conditions $\left[D\frac{\partial C(x, t | x_{0})}{\partial x} + \frac{U_0}{\zeta x}C(x, t | x_{0})\right]_{x=0} = 0$ and $C(x \to \infty, t  | x_{0} ) = 0$. Laplace transforming Eq. (\ref{log1}) we obtain
\begin{align}\label{G:1}
   \frac{d^{2} \tilde{C}(x, s \mid x_0)}{d x^{2}}  +& \frac{d}{d x}\Big[\left(\frac{U_{0}}{D\zeta x}\right) \tilde{C}(x, s \mid x_0)\Big]\\
   &- \frac{s}{D}\tilde{C}(x, s \mid  x_0) + \frac{\delta(x-x_0)}{D}=0 ,  \nonumber
\end{align}
which is a second order, inhomogeneous ordinary differential equation. 

Following Ray and Reuveni \cite{ray2020diffusion}, we start by employing the change of variables $\tilde{C}(x, s \mid x_0) = x^n\tilde{c}(x, s \mid x_0)$, where $n = 1/2 - U_0/2 D \zeta \equiv 1 - \nu$ and we note that $(n-1)(n+U_0/D \zeta)= -\nu^2 $.  For every $x \not = x_0$, Eq. (\ref{G:1}) then reads
\begin{align}\label{G:2}
   \frac{d^{2} \tilde{c}(x, s \mid x_0)}{d x^{2}}  + & \frac{1}{x}\frac{d \tilde{c}(x, s \mid x_0)}{d x}
  \\
  & - \left(\frac{\nu^2}{x^2} + \frac{s}{D}\right)\tilde{c}(x, s \mid  x_0) =0 ,  \nonumber
\end{align}
which is a modified Bessel's equation. The general solution for $\tilde{c}(x, s \mid  x_0)$ is 
\begin{equation} \label{G:3}
\tilde{c}(x, s \mid x_0) = \begin{cases}A_{1}(s) I_{ \nu}\left(\sqrt{\frac{s}{D}} x\right)+B_{1}(s) K_{ \nu}\left(\sqrt{\frac{s}{D}} x\right) ,& x > x_{0} \\ A_{2}(s) I_{ \nu}\left(\sqrt{\frac{s}{D}} x\right)+B_{2}(s) K_{\nu}\left(\sqrt{\frac{s}{D}} x\right) ,& x<x_{0}\end{cases},
\end{equation}
where $I_{\nu}(\cdot)$ and $K_{\nu}(\cdot)$ are modified Bessel functions of the first
and second kind of order $ \nu$, respectively. Recalling that $\tilde{C}(x, s \mid x_0) = x^n\tilde{c}(x, s \mid x_0)$, the general solution for $\tilde{C}(x, s \mid  x_0)$ is
\begin{align} \label{G:4}
\tilde{C}(x, s \mid x_0) = & x^{1-\nu} \times \\
&\begin{cases}A_{1}(s) I_{ \nu}\left(\sqrt{\frac{s}{D}} x\right)+B_{1}(s) K_{ \nu}\left(\sqrt{\frac{s}{D}} x\right) ,& x > x_{0} \\ A_{2}(s) I_{ \nu}\left(\sqrt{\frac{s}{D}} x\right)+B_{2}(s) K_{ \nu}\left(\sqrt{\frac{s}{D}} x\right) ,& x<x_{0}\end{cases}. \nonumber
\end{align}

To determine the linear combination coefficients $A_i(s)$ and $B_i(s)$ we consider the initial and boundary conditions. First, the boundary condition $\tilde{C}(x \to \infty, s  | x_{0} ) = 0$ requires that $A_1 = 0$, since in the limit $x \to \infty$ we have $I_{\nu}(x) \to \infty$ and $K_{\nu}(x) \to 0$). The other boundary condition is $\left[D\frac{d \tilde{C}(x, s | x_{0})}{dx} + \frac{U_0}{\zeta x}\tilde{C}(x, s | x_{0}) \right]_{x=0} = 0$. Imposing this condition is not straightforward, first we note that since $U_0 + \zeta D (1-2 \nu) = 0$ we have
\begin{align} \label{G:5}
&\left[D\frac{d \tilde{C}(x, s | x_{0})}{dx} + \frac{U_0}{\zeta x}\tilde{C}(x, s | x_{0}) \right]_{x=\epsilon} = 
\\
&\frac{\sqrt{D s}}{\epsilon^{v-1}}\left[A_{2} I_{v-1} \left(\sqrt{\frac{s}{d}} \epsilon\right)-B_{2} K_{v-1} \left(\sqrt{\frac{s}{d}} \epsilon\right)\right] . \nonumber
\end{align}

\noindent As $\epsilon \to 0$, we take the limiting forms of $I_{\nu}$ and $K_{\nu}$ and impose that this quantitiy must be equal to zero. We then get that for all $\nu < 1$
\begin{equation} \label{G:6}
    B_2(s) = \frac{2 \operatorname{Sin}(\pi v) }{\pi}A_2(s).
\end{equation}

Next, Eq. (\ref{G:4}) must be continuous at $x_0$, which translates to 
\begin{equation}\label{G:7}
 B_{1}(s)= A_{2}(s)\frac{ I_{ \nu}\left(\sqrt{\frac{s}{D}} x_0\right)}{ K_{\nu}\left(\sqrt{\frac{s}{D}} x_0\right)} + B_2(s) .
\end{equation}

\noindent The last condition comes from integrating Eq. (\ref{G:1}) over a very small interval $[x_0-\Delta, x_0+\Delta]$, which gives
\begin{align}\label{G:8}
   &\left[\frac{d \tilde{C}(x, s \mid x_0)}{d x}\right]^{x_0+\Delta}_{x_0-\Delta} + \Big[\left(\frac{U_{0}}{D\zeta x}\right)  C(x, t \mid x_0)\Big]^{x_0+\Delta}_{x_0-\Delta}\\
   &- \frac{s}{D} \int^{x_0+\Delta}_{x_0-\Delta}\tilde{C}(x, s \mid  x_0) dx + \frac{1}{D} \int^{x_0+\Delta}_{x_0-\Delta}\delta(x-x_0) dx =0 .   \nonumber
\end{align}

\noindent The imposed continuity of $\tilde{C}(x, s \mid x_0)$ at $x_0$ means that as we take $\Delta \to 0$ the two middle terms go to zero. The last term goes to $1/D$ (by definition of the delta function), and we are left with the condition 
\begin{equation}\label{G:9}
 \lim_{\Delta \to 0} \left[\frac{d \tilde{C}(x, s \mid x_0)}{d x}\right]^{x_0+\Delta}_{x_0-\Delta}=-\frac{1}{D}.  
\end{equation}

\noindent Plugging Eq. (\ref{G:4}) into Eq. (\ref{G:9}) we obtain
\begin{align}\label{G:10}
  & (B_2(s) -B_1(s))  \left[  K_{\nu} \left(\sqrt{\frac{s}{D}} x_0\right) - \sqrt{\frac{s}{D}} x_0 K_{\nu+1} \left(\sqrt{\frac{s}{D}} x_0\right)  \right]  \nonumber
   \\
   &-A_2(s) \left[ I_{\nu} \left(\sqrt{\frac{s}{D}} x_0\right) + \sqrt{\frac{s}{D}} x_0 I_{\nu+1} \left(\sqrt{\frac{s}{D}} x_0\right)  \right]
   \\
   &=-\frac{ x_0^\nu}{D}  \nonumber
\end{align} 

Equations  (\ref{G:6}), (\ref{G:7}) and Eq. (\ref{G:10}) constitute a system of linear equations.  Solving, we obtain
\begin{equation}\label{G:11}
\begin{array}{ll}
    B_1(s) =  \frac{x_0^\nu}{D} I_{-\nu} \left(\sqrt{\frac{s}{D}} x_0\right), \\ \\
    A_2(s) = \frac{x_0^\nu}{D} K_{\nu} \left(\sqrt{\frac{s}{D}} x_0\right) , \\ \\
    B_2(s) = A_2(s) \frac{2 \operatorname{Sin}(\pi \nu)}{\pi} 
\end{array}
\end{equation}
The conserved propagator is thus (note that $\nu < 1$)
 
\begin{align}\label{G:12}
&\tilde{C}(x, s \mid x_0) = \frac{x}{D}\left(\frac{x_0}{x}\right)^\nu \times
\\
& \begin{cases}  I_{-\nu} \left(\sqrt{\frac{s}{D}} x_0\right) K_{ \nu}\left(\sqrt{\frac{s}{D}} x\right) ,& x > x_{0} \\  K_{\nu} \left(\sqrt{\frac{s}{D}} x_0\right) \left[I_{\nu}\left(\sqrt{\frac{s}{D}} x\right) + \frac{2 \operatorname{Sin}(\pi \nu)}{\pi} K_{\nu}\left(\sqrt{\frac{s}{D}} x\right) \right],& x<x_{0}\end{cases}. \nonumber
\end{align}

\subsection{First-passage time to a point $b$, different than the origin}

While the distribution of the first-passage time to the origin for a particle diffusing in a logarithmic potential is known \cite{bray2000random}, it is crucial to realize that under the logarithmic potential, the first-passage time depends explicitly on the target location $b$, and not just on $\abs{b-x_0}$, i.e., the distance to it. For our calculation, we require the distribution of the first-passage time to the target upper boundary $b > 0$, which is rederived in this appendix for completeness. 

To compute the first-passage distribution we first write the differential equation for the Green’s function of diffusion in logarithmic potential (Eq. (\ref{log1})), where we require the initial condition $C(x, t=0 | x_{0}>b)=\delta\left(x-x_{0}\right)$ and boundary conditions $C(b, t | x_{0}) = 0$ and $C(x \to \infty, t  | x_{0} ) = 0$.

Following Ray and Reuveni \cite{ray2020diffusion} we note that Eq. (\ref{log1}) is a Forward Fokker-Planck equation for which the corresponding backward Fokker-Planck equation is \cite{paul1999stochastic}
\begin{equation}\label{G:13}
        \frac{\partial  C(x, t \mid  x_0)}{\partial t}  =   -\frac{U_{0}}{\zeta x_0}\frac{\partial C(x, t \mid x_0)}{\partial x_0}+D \frac{\partial^{2} C(x, t \mid x_0)}{\partial x_0^{2}}. 
\end{equation}
By integrating out $x$ in Eq. (\ref{G:13}) over the entire domain we get the corresponding partial differential equation for the survival probability
\begin{equation}\label{G:14}
        \frac{\partial  S(t \mid  x_0)}{\partial t}  =   -\frac{U_{0}}{\zeta x_0}\frac{\partial S(t \mid x_0)}{\partial x_0}+D \frac{\partial^{2} S(t \mid x_0)}{\partial x_0^{2}}. 
\end{equation}

The initial condition $C(x, t=0 | x_{0}>b)=\delta\left(x-x_{0}\right)$ corresponds to $S(t=0\mid x_0>b) = 1$. Thus, when Laplace transforming Eq. (\ref{G:14}) we obtain
\begin{equation}\label{G:15}
  \frac{d^{2} \tilde{S}(s \mid x_0)}{d x_0^{2}}  -\frac{U_{0}}{D \zeta x_0}\frac{d \tilde{S}(s \mid x_0)}{d x_0} - \frac{s}{D}\tilde{S}(s \mid  x_0) = -\frac{1}{D} ,  
\end{equation}
which is a second order, inhomogeneous ordinary differential equation. To convert it to a homogeneous equation we consider the transformation $\tilde{S}(s \mid x_0) = \tilde{z}(s \mid x_0) + \frac{1}{s}$, which yields

\begin{equation}\label{G:16}
  \frac{d^{2} \tilde{z}(s \mid x_0)}{d x_0^{2}}  -\frac{U_{0}}{D \zeta x_0}\frac{d \tilde{z}(s \mid x_0)}{d x_0} - \frac{s}{D}\tilde{z}(s \mid  x_0) = 0.  
\end{equation}

To solve, we employ the change of variables $\tilde{z}(s \mid x_0) = x_0^\nu\tilde{y}(s \mid x_0)$, where $\nu = 1/2 + U_0/2 D \zeta$ and note that $\nu(\nu-1) - U_0 \nu/D \zeta = -\nu^2 $. Equation (\ref{G:16}) then reads
\begin{align}\label{G:17}
  x^2_0 \frac{d^{2} \tilde{y}(s \mid x_0)}{d x_0^{2}}  + & x_0\frac{d \tilde{y}(s \mid x_0)}{d x_0}
  \\
  & - \left(\frac{s}{D} x^2_0 + \nu^2 \right)\tilde{y}(s \mid  x_0) = 0 ,  \nonumber
\end{align}
which is a modified Bessel's equation. The general solution is 
\begin{equation} \label{G:18}
\tilde{y}(s \mid x_0) = A(s) I_{\nu}\left(\sqrt{\frac{s}{D}} x_{0}\right)+B(s) K_{\nu}\left(\sqrt{\frac{s}{D}} x_{0}\right),
\end{equation}
where $I_{\nu}(\cdot)$ and $K_{\nu}(\cdot)$ are modified Bessel functions of the first
and second kinds of order $\nu$, respectively. Recalling that $\tilde{z}(s \mid x_0) = x_0^\nu\tilde{y}(s \mid x_0)$ and that $\tilde{S}(s \mid x_0) = \tilde{z}(s \mid x_0) + \frac{1}{s}$ , the general solution is 
\begin{equation} \label{G:19}
\tilde{S}(s \mid x_0) = A(s) x_0^\nu I_{\nu}\left(\sqrt{\frac{s}{D}} x_{0}\right)+B(s) x_0^\nu K_{\nu}\left(\sqrt{\frac{s}{D}} x_{0}\right) + \frac{1}{s}.
\end{equation}

To determine the combination coefficients $A(s)$ and $B(s)$ we consider the initial and boundary conditions. In the limit $x_0 \to \infty$ we have $S(t \mid x_0) \to 1$, hence $\tilde{S}(s \mid x_0) \to \frac{1}{s}$. Taking this limit in Eq. (\ref{G:19}) we obtain the condition $A(s) x_0^\nu I_{\nu}\left(\alpha x_{0}\right)+B(s) x_0^\nu K_{\nu}\left(\alpha x_{0}\right) =0 $. As $x_0 \to \infty$, $I_{\nu}(x_0) \to \infty$ and $K_{\nu}(x_0) \to 0$, and so for the condition to hold we must set $A(s)=0$. This gives
\begin{equation} \label{G:20}
\tilde{S}(s \mid x_0) = B(s) x_0^\nu K_{\nu}\left(\sqrt{\frac{s}{D}} x_{0}\right) + \frac{1}{s}.
\end{equation}

The other boundary condition is an absorbing boundary condition at $b$. In terms of the survival probability it states $\tilde{S}(s \mid b)=0$. By imposing this condition on Eq. (\ref{G:20})  we obtain 
\begin{equation} \label{G:21}
B(s) = - \frac{1}{s b^\nu K_{\nu}\left(\sqrt{\frac{s}{D}} b\right)}
\end{equation}
Thus, the Laplace transform of the survival probability is
\begin{equation} \label{G:22}
\tilde{S}(s \mid x_0) =  \frac{1}{s} \left[ 1 - \left(\frac{x_{0}}{b}\right)^\nu\frac{K_{\nu}\left(\sqrt{\frac{s}{D}} x_{0}\right)}{ K_{\nu}\left(\sqrt{\frac{s}{D}} b\right)}  \right].
\end{equation}

\noindent Recall that $S(t \mid x_0)=1-\int_{0}^{t} f\left(t^{\prime} \mid x_0\right) d t^{\prime}$, where $f\left(t \mid x_0\right)$ is the first-detection distribution of a particle with initial state $x_0$. It is thus apparent that its Laplace transform is (note $b>0$)
\begin{equation} \label{G:23}
\tilde{T}_f(b,s \mid x_0)=\left(\frac{x_{0}}{b}\right)^\nu\frac{K_{\nu}\left(\sqrt{\frac{s}{D}} x_{0}\right)}{ K_{\nu}\left(\sqrt{\frac{s}{D}} b\right)}  .
\end{equation}


\end{document}